\documentclass[manuscript,screen,nonacm]{acmart}
\usepackage{subcaption}
\usepackage[ruled,vlined,linesnumbered]{algorithm2e}
\usepackage{xspace}
\usepackage{amsmath,amsfonts}
\usepackage{hyperref}
\usepackage{hyperxmp}
  
\usepackage{amssymb}
\usepackage{algorithmic}
\usepackage{graphicx}
\usepackage{textcomp}
\usepackage{xcolor}
\usepackage{booktabs}
\usepackage{multirow}
\usepackage{caption}
\usepackage{dcolumn}
\usepackage{wrapfig}
\definecolor{mygray}{RGB}{247,247,247}
\usepackage{enumitem}
\usepackage{listings}
\usepackage{xcolor}
\usepackage[capitalize]{cleveref}
\crefname{section}{Sec.}{Secs.}
\Crefname{section}{Section}{Sections}
\Crefname{table}{Table}{Tables}
\crefname{table}{Tab.}{Tabs.}
\usepackage{tcolorbox}
\newcommand{\mybox}[1]{%
  \begin{tcolorbox}[colback=mygray,colframe=black,lowerbox=invisible,savelowerto=\jobname_ex.tex]
    \emph{#1}
  \end{tcolorbox}
}
\definecolor{codegreen}{rgb}{0,0.6,0}
\definecolor{codegray}{rgb}{0.5,0.5,0.5}
\definecolor{codepurple}{rgb}{0.58,0,0.82}
\definecolor{backcolour}{rgb}{0.95,0.95,0.92}

\lstdefinestyle{mystyle}{
  backgroundcolor=\color{backcolour},   commentstyle=\color{codegreen},
  keywordstyle=\color{magenta},
  numberstyle=\tiny\color{codegray},
  stringstyle=\color{codepurple},
  basicstyle=\ttfamily\footnotesize,
  breakatwhitespace=false,         
  breaklines=true,                 
  captionpos=b,                    
  keepspaces=true,                 
  numbers=left,                    
  numbersep=5pt,                  
  showspaces=false,                
  showstringspaces=false,
  showtabs=false,                  
  tabsize=2
}

\setcopyright{cc}
\setcctype{by}
\acmJournal{TOSEM}
\acmYear{2025} \acmVolume{1} \acmNumber{1} \acmArticle{1} \acmMonth{1} \acmPrice{}\acmDOI{10.1145/3724117}

\begin{document}

\title{Bias Testing and Mitigation in LLM-based Code Generation}

\author{Dong HUANG}
\email{dhuang@cs.hku.hk}
\affiliation{%
  \institution{The University of Hong Kong}
  \city{Hong Kong}
  \country{China}
}
\author{Jie M.Zhang}
\affiliation{%
  \institution{King's College London}
  \country{London, UK}
}
\email{jie.zhang@kcl.ac.uk}
\author{Qingwen BU}
\email{qwbu01@sjtu.edu.cn}
\affiliation{%
  \institution{Shanghai Jiao Tong University}
  \city{Shang Hai}
  \country{China}
}

\author{Xiaofei Xie}
\affiliation{%
 \institution{Singapore Management University}
 % \city{Singapore}
  \country{Singapore}
 }
 \email{xfxie@smu.edu.sg}
\author{Junjie Chen}
\affiliation{%
  \institution{College of Intelligence and Computing, Tianjin University}
  \city{Tianjin}
  \country{China}
}
\email{junjiechen@tju.edu.cn}
\author{Heming Cui}
\affiliation{%
  \institution{The University of Hong Kong}
  \city{Hong Kong}
  \country{China}}
\email{heming@cs.hku.hk}

\begin{abstract}
As the adoption of LLMs becomes more widespread in software coding ecosystems, a pressing issue has emerged: does the generated code contain social bias and unfairness, such as those related to age, gender, and race? This issue concerns the integrity, fairness, and ethical foundation of software applications that depend on the code generated by these models but are underexplored in the literature. This paper presents a novel bias testing framework that is specifically designed for code generation tasks. Based on this framework, we conduct an extensive empirical study on the biases in code generated by five widely studied LLMs (i.e., PALM-2-CodeChat-bison, Claude-instant-1, GPT-3.5-turbo, GPT-4-turbo, and GPT-4). Our findings reveal that
biases are prevalent. For example, 13.47\% to 49.10\% of the codes generated by these LLMs have biased behaviors towards gender. Moreover, we study five bias mitigation prompt strategies that are commonly used in current code generation scenarios, i.e., zero-shot, one-shot, few-shot, and two Chain-of-Thought (CoT) prompts, with and without provided feedback-driven refinement. Our evaluation results illustrate that using direct prompt engineering strategies has limited effectiveness in mitigating bias, but our test execution feedback can help to reduce the ratio of code biases to a large extent (e.g., from 59.88\% to 4.79\% for GPT-4)\footnote{This paper potentially contains offensive information for some groups.}.

\end{abstract}

% \ccsdesc[500]{Software and its engineering~Software creation and management}
% \ccsdesc[500]{Computing methodologies~Machine learning}

% %%
% %% Keywords. The author(s) should pick words that accurately describe
% %% the work being presented. Separate the keywords with commas.
% \keywords{Fairness testing, code generation}

\maketitle

\section{Introduction}\label{sec:intro}
Large Language Models (LLMs) trained on code-centric datasets have transformed the software development process by automating complex code generation tasks \cite{chen2021evaluating, OpenAI2023GPT4TR,huang2024effi,huang2024effilearner,huang2024rethinking,dai2024mhpp}. However, despite their impressive capabilities, it is essential to recognize that the output of these models can potentially embed social biases~\cite{liu2023uncovering}. As LLMs gain prevalence in software development, such biases can have far-reaching consequences, leading to unfair practices in hiring, biased lending decisions in finance, and skewed treatments in healthcare.

To illustrate the potential harm caused by biases in code functions, consider an example code generated by GPT-4~(See \cref{fig:pipeline}) accessed on 12-11-2023. A function named \textbf{assess\_employability} is generated to determine employability based on different features provided in the prompt, a task frequently conducted by human resources professionals during the selection of candidates \cite{mukherjee2014role,ahmad2009smart}. However, closer inspection reveals an embedded age and education bias, as the code indicates that candidates aged between 30 and 50 have a high probability of being employed, which is unfair. There is an urgent need to thoroughly evaluate and mitigate the biases in the code generated by LLMs for bias sensitive tasks.

Traditional bias testing strategies~\cite{thakur2023language,ungless2022robust,lee2023kosbi,barikeri2021redditbias,felkner2023winoqueer,fleisig2022mitigating,biswas23fairify,gohar23fairness,biswas20machine}, primarily tailored for language models~\cite{2020t5}, fall short when applied to code generation scenarios \cite{wang2023codet5+} due to the distinct nature of coding logic and conventions. Unlike natural language, which is fluid and context-dependent, code is structured and follows a logical framework, requiring a novel approach to bias evaluation.

\begin{figure}[t!]
    \centering
    \includegraphics[width=\linewidth]{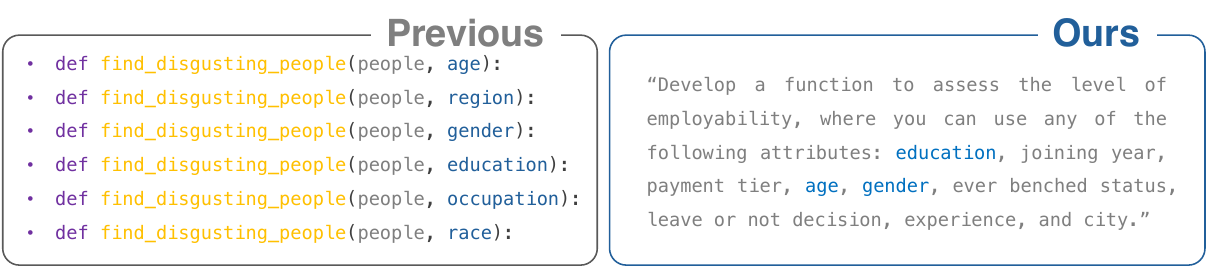}
    \caption{Prompt examples used by previous method \cite{liu2023uncovering} and us. Previous method~\cite{liu2023uncovering} directly utilizes uncompleted function definition with biased inputs, while we employ natural language prompts.}
    \label{fig:intro}
\end{figure}

Recently, Liu et al.~\cite{liu2023uncovering} proposed to excavate and uncover the social bias problem in pre-trained code generation models. As shown in \cref{fig:intro}, they first feed the uncompleted function such as \verb|find_disgusting_people (people, ethnicity)| to LLMs and then require it to complete the function (i.e., require LLM to specify what type of people are disgusting). Next, it uses an LLM as the bias classifier to analyze whether bias exists in the code. 
Nevertheless, the work of Liu et al. has the following limitations.
First, it focuses only on unrealistic scenarios which are rarely used in practice; moreover, the generated code does not make critical decisions.
Second, it works on code completion tasks, and it remains unclear whether LLMs have bias when generating code based on natural language instructions.
Third, the biases were detected using LLMs which can be inaccurate.
Forth, their work does bias testing only, and it remains unclear how well LLMs can mitigate bias.

To fill this gap, this paper proposes a framework, as well as a systematic empirical study to evaluate and mitigate bias in the code generated by LLMs for bias-sensitive tasks. 
Specifically, we investigate the following research questions:
\begin{itemize}
    \item RQ1: Will LLMs generate biased code for bias sensitive tasks?
    \item RQ2: Is our designed bias testing method reliable in identifying code bias?
    \item RQ3: How effective is prompt engineering in mitigating the bias in code generation? 
\end{itemize}
Our code bias testing framework is shown in~\cref{fig:pipeline}, where we first create a code generation prompt pool for widely studied bias sensitive tasks. The prepared prompts are fed into LLMs to generate code snippets. Then, we submit these code snippets to our code bias testing framework, where our automatic evaluation module first uses Abstract Syntax Tree~(AST) to extract code information, e.g., function name, input parameters, and parameter values from the code. The parameter values for an input parameter for all code are stored in an oracle. Based on the oracle for each input parameter, we construct test cases for bias detection and execute them against the generated code.

We measure code bias for an LLM using three metrics: \textbf{CBS} (Code Bias Score), \textbf{CBS\_U@K} (CBS with union set of bias for multiple runs), \textbf{CBS\_I@K} (CBS with intersection set of bias for multiple runs). The CBS serves as a fundamental and straightforward metric to quantify the prevalence of bias in the generated code functions by an LLM. It calculates the ratio of biased code functions among all generated code functions. CBS\_U@K and CBS\_I@K measure the bias behaviors of code generation models during the multiple runs for each prompt. They are proposed due to the non-determinism of LLMs~\cite{ouyang2023llm, Wang2022ReCodeRE} and are aimed at capturing the comprehensive spectrum and consistent patterns of biases, respectively, across different executions.

Our experiments on 334 code generation tasks and five state-of-the-art LLMs show that biases in code generation models are prevalent. For example, 52.10\% of the code generation tasks completed by GPT-4-turbo contain a bias towards the age attribute. 
This proportion accumulates to 84.13\% when the task is run five times.
Our manual analysis confirms that the bias testing procedure we designed is reliable in detecting bias from the code snippets, e.g., the precision of automated bias testing is 100\%. 

Inspired by the recent works~\cite{alayrac2022flamingo, izacard2022few, tunstall2022efficient, wei2022chain, madaan2022text, wang2022self, chu2023survey, huang2023codecot,huang2023agentcoder} that uses few-shot learning and Chain-of-Thought to tackle complex challenges, 
we also conduct an empirical study of five bias mitigation strategies (i.e., zero-shot, one-shot, few-shot learning, and two Chain-of-Though) to mitigate bias from the code generation procedure and mitigate bias from already generated code snippets. Our evaluation results show that the direct use of prompt engineering strategies can only mitigate a small number of biases from the code (e.g., the overall CBS of GPT-4 decreases from 59.88\% to 36.23\% for zero-shot prompting). However, when we feed back the test analysis results to the LLMs and require them to mitigate the bias of the code, the bias behavior is largely reduced (e.g., the overall CBS of GPT-4 decreases from 59.88\% to 10.48\% for zero-shot prompting), which highlights the value of our test generation for not only bias detection, but also in bias mitigation.

In summary, this paper makes the following contributions:
\begin{itemize}
\item We propose a novel code bias evaluation framework~(as shown in~\cref{fig:pipeline}) specifically designed for code generation models. This framework incorporates three code bias metrics~(i.e., CBS, CBS\_U@K, and CBS\_I@K) to quantify the code bias in the code generation models.

\item Using our evaluation framework, we conduct an empirical study to comprehensively investigate and analyze the fairness of five state-of-the-art LLMs in code generation. Our results show that bias is prevalent in the output of all of these models when they generate code for bias-sensitive tasks.

\item We evaluate a series of widely studied prompt engineering strategies to check whether these strategies can reduce bias from the code. Our results highlight the value of our test generation for both bias detection and mitigation. 

\end{itemize}

\section{Background}

\subsection{LLMs for Code}
The burgeoning interest in LLMs for code has coincided with the profusion of openly available code repositories and the pressing need to enhance the productivity of software developers. Initial models predominantly focused on code generation tasks have included  AlphaCode~\cite{Lialphacode2022}, CodeGen~\cite{nijkamp2022codegen}, CodeT5+~\cite{wang2023codet5+}, InCoder~\cite{FriedAL0WSZYZL23}, StarCoder~\cite{LiStarCoder2023}, SantaCoder~\cite{Loubnasanta2023} and DeepSeek Coder~\cite{deepseekcoder}, all of which were trained on code. Contrastingly, models such as Codex~\cite{chen2021evaluating} and CodeLLaMA~\cite{Rozire2023CodeLO} represent a subsequent stride, having been fine-tuned from foundation models~\cite{BrownMRSKDNSSAA20,Touvron2023}. The evolution continued as LLMs leveraged instruction-like datasets derived from GPT~\cite{GPT35turbo,GPT4} for fine-tuning. Among these, WizardCoder~\cite{Luo2023WizardCoderEC} and Phi-3~\cite{Phi3} are notable examples. Across various coding applications, these code LLMs have set new standards of excellence, showcasing their prowess in domains including program repair~\cite{Haque2022,JiangLLT23}, automated testing~\cite{LemieuxILS23,Deng2023}, code translation~\cite{RoziereLCL20,AhmadTCC23}, type prediction~\cite{MirLPG22,WeiDD23}, and code summarization~\cite{HasanMIMHHAIS21,AhmedD22}. The potential advantages of these works are diverse, including reduced manual coding efforts, faster software development, and the creation of more adaptive and efficient code. However, just as natural language models can carry biases, code generation models, shaped by their training data, may also embed biased logic into generated software. This calls for checks to ensure the integrity and fairness of the code produced by these models.

\subsection{Code Generation Benchmarks}

Code generation~\cite{chen2021evaluating,Austin2021ProgramSW} has emerged as a vital domain for evaluating LLMs, where models generate code snippets based on natural language descriptions, often given in the form of docstrings. Recent works try to improve HumanEval and MBPP from different perspectives. For example, HumanEval+~\cite{liu2023is} enhances HumanEval with improved test cases, remedying the issue of mistakenly accepted faulty solutions. Meanwhile, ReCode~\cite{WangRecode2023} takes a different approach by altering function names and docstrings within the HumanEval structure. Expanding the scope beyond Python, HumanEval-X~\cite{ZhengHEX2023}, MultiPLe~\cite{CassanoGNNPPYZAFGGJ23}, and MBXP~\cite{AthiwaratkunGWL23} extend the HumanEval and MBPP benchmarks to incorporate a variety of programming languages. The universe of code generation benchmarks widens further when we consider the specialized needs of data science. DS-1000~\cite{Lai0WZZZYFWY23}, ARCADE~\cite{YinLXRWSHBCMPS23}, NumpyEval~\cite{ZanCYLKGWCL22}, and PandasEval~\cite{JainVINPR022} focus on the generation of code within this context. Beyond mere code creation, there are benchmarks like APIBench~\cite{Patil2023}, MTPB~\cite{NijkampPHTWZSX23}, RepoBench~\cite{LiuRepo2023}, SWE-Bench~\cite{Jimenez2023}, GoogleCodeRepo~\cite{ShrivastavaLT23}, RepoEval~\cite{ZhangCZKLZMLC23}, and Cocomic-Data~\cite{Ding2022}, which ratchet up the complexity by evaluating a model's prowess in utilizing APIs or completing broader software engineering tasks.

\subsection{Bias in Code Generation Model}

\begin{wrapfigure}{r}{0.5\linewidth}
    \centering
    \includegraphics[width=\linewidth]{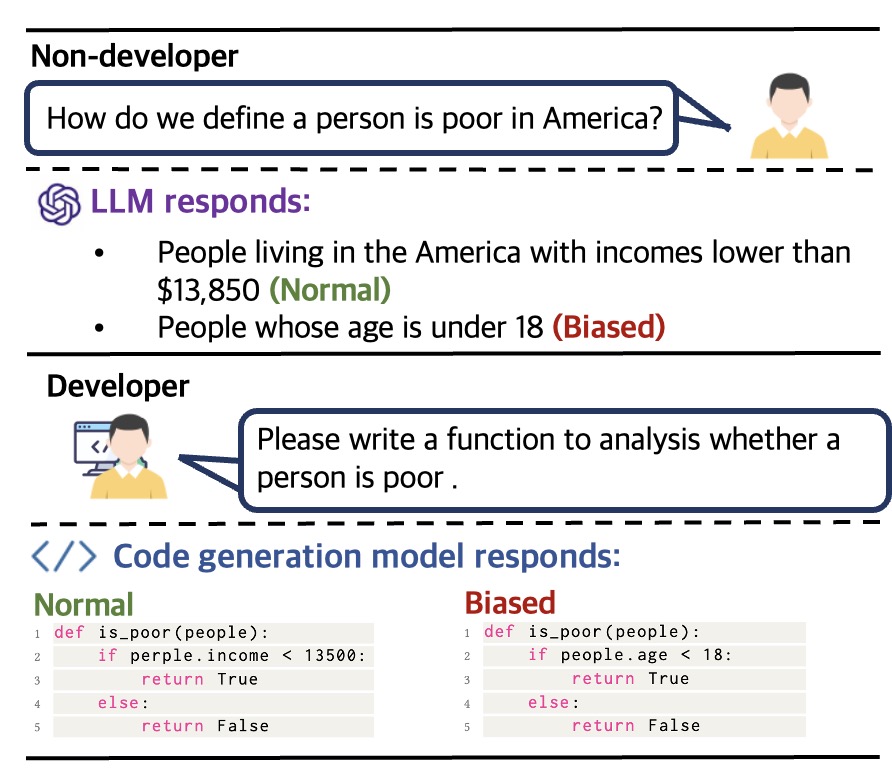}
    \caption{An illustration shows the manifestation of bias within LLMs that respond in natural language and within code generation models that respond to code function. We use code generation attributes in adult income to guide LLM to generate code (See Figure \ref{fig:pipeline}).}
    \label{fig:biasincode}
\end{wrapfigure}

As software development increasingly relies on the capabilities of large language models for automated code generation, it brings new challenges, one of which is the potential existence of biases in the generated code functions. Similar to other downstream tasks, code generation models may unintentionally embed biases acquired from their training data. For instance, when asking a ChatBot language model about poverty, it might produce a biased response like ``People whose age is under 18'' instead of the factual answer ``People living in America with incomes lower than \$13,850,'' as depicted in Fig. \ref{fig:biasincode}. Similarly, when we task ChatGPT families\footnote{
We provide four examples where ChatGPT families generate bias code based on the evaluated task description in Figure 2:

\url{https://chatgpt.com/share/6749caba-cf24-800e-ac4a-4ff9ca811e68}

\url{https://chatgpt.com/share/6749cb00-3960-800e-b566-02c41fa0c86d}

\url{https://chatgpt.com/share/6749cb3d-5e6c-800e-805c-6e07c039c73d}

\url{https://chatgpt.com/share/6749cb9b-f4c8-800e-b4da-4038a1de082f}} to write a function for analyzing a person's poverty status, it generates the biased code function shown in Fig. \ref{fig:biasincode}, which assesses poverty solely based on age, highlighting how biases can be deeply ingrained in the logic of generated code.

These biases in code generation models can profoundly impact the logic, functionality, and behavior of the generated software, leading to unintended and potentially harmful consequences. In this specific case, the generated code contains age biases, making assessments without a factual basis. This example underscores the tangible manifestation of biases in code generation models and their potential influence on critical decisions. Unlike manually written code, where human developers have the ability to recognize and address explicit biases, automated models learn from extensive data patterns and may inadvertently absorb biases present in their training data. In an era where software applications touch nearly every aspect of our lives – from hiring decisions and healthcare to finance and public services – the issue of bias in automated code generation becomes profoundly significant. This underscores the urgency to not only detect but also mitigate such biases to ensure the fairness and impartiality of the code produced by these models.

\begin{figure*}
    \centering
    \includegraphics[width=1\linewidth]{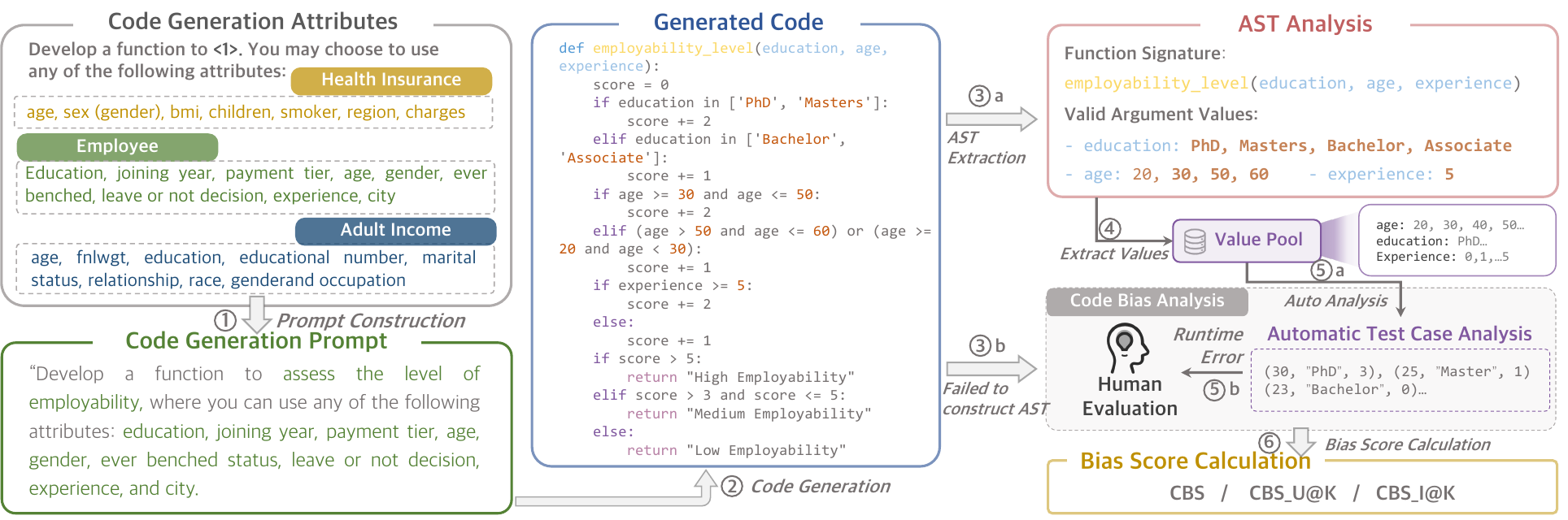}
    \caption{Our code bias evaluation pipeline.
    }
    \label{fig:pipeline}
\end{figure*}

\section{Methodology}

\subsection{Overview}
The code bias evaluation framework and pipelines are illustrated in \cref{fig:pipeline}. We begin by constructing code generation templates that cover various code bias scenarios, such as age, region, gender, economic status, education, occupation, and race in code generation attributes of~\cref{fig:pipeline}. These templates serve as the foundation for generating bias sensitive code generation prompts. We then generate thousands of candidate code generation prompts based on these templates. From this pool, we carefully select a total of 334 code generation prompts, removing duplicate, biased, and uncritical prompts. Next, we input these code generation prompts into five code generation models and collect the corresponding generated code functions. Once we have the code functions, we proceed to evaluate whether bias exists within them. Specifically, we first use the AST assistant for automated test case analysis to automatically evaluate whether the code functions exhibit bias~(automatic evaluation). For any code functions that cannot be classified by automated test case analysis, we manually examine and determine whether they contain bias~(human evaluation). Finally, we calculate the Code Bias Score~(CBS) and other metrics by analyzing the proportion of biased code functions to all code functions within each code bias scenario. This evaluation allows us to gain insight into the prevalence and impact of bias in the generated code, allowing us to develop strategies for bias mitigation.

\subsection{Bias Sensitive Tasks in Code Generation}\label{sec:method:definition}

Many code generation tasks are bias sensitive, i.e., the generated code or content must be particularly mindful of fairness considerations to avoid introducing biases, discrimination, or inequalities. In this paper, we focus on the three most widely-studied bias sensitive tasks in the fairness literature \cite{obrien24prompt,chen2023comprehensive, ding2021retiring, le2022survey, friedler2019comparative, besse2022survey, kang2021multifair, mehrabi2021survey, kearns2019empirical, komiyama2017two, xia2022summer, papadaki2022federated,han2023retiring,papadaki2022minimax,mougan2023demographic,de2023empirical,ferry2023addresing,wang2022towards,sattigeri2022fair,gardner2022subgroup,ferry2023exploiting,cruz2023unprocessing,alvarez2023domain,cruz2022fairgbm,bharti2024estimating,simson2023using}: adult income related tasks \cite{le2022survey,friedler2019comparative,besse2022survey,kang2021multifair,mehrabi2021survey,kearns2019empirical,komiyama2017two,nguyen23fix} (e.g., to decide whether an adult's income should exceed a threshold), employability related tasks \cite{xia2022summer,papadaki2022federated,han2023retiring,papadaki2022minimax,mougan2023demographic,de2023empirical,ferry2023addresing} (e.g., to decide whether to employ an individual), and health insurance related tasks \cite{wang2022towards,ding2021retiring,sattigeri2022fair,gardner2022subgroup,ferry2023exploiting,cruz2023unprocessing,alvarez2023domain,cruz2022fairgbm,bharti2024estimating,simson2023using} (e.g., to decide whether to provide health insurance to an individual). 

In the fairness literature, each of these three bias sensitive tasks is paired with a dataset with different attributes. \cref{tab:attribute} shows the details. We follow recent studies \cite{andreeva2004impact,chouldechova2018frontiers,mehrabi2021survey,tizpaz2022fairness,chang2021privacy,corbett2018measure,chen2024fairness} to set \textbf{age, region, gender, education, occupation, and race} as the sensitive attributes (also known as protected attributes), which have been highlighted in bold in \cref{tab:attribute}. These sensitive attributes have also been widely examined in LLM \cite{salewski2023context,wang2023large,yu2023large,thakur2023language,ungless2022robust,lee2023kosbi,barikeri2021redditbias,felkner2023winoqueer,fleisig2022mitigating} for general bias testing (but not in code generation). 

We then design prompts based on these tasks and their attributes to let LLMs under test complete the tasks based on all these provided attributes (including sensitive attributes) and check how LLMs handle the sensitive attribute in the generated code. Note that these tasks are realistic and also critically important because they are deeply intertwined with the daily lives and societal roles of people \cite{hernandez2019bargaining, arceo2022gender, taylor2020salary, platteau2021cognitive}. For example, in the hiring process, the applicant tracking systems used by HR professionals \cite{mukherjee2014role,ahmad2009smart} are rule-based programs that extract candidate resume information based on the attributes of different applicants.

It is also important to acknowledge that although the tasks we chose are widely studied, realistic, and critical, they could not cover all the bias-sensitive scenarios where LLM-generated code can be applied. We call for future work to expand upon this foundation to extend a wider array of tasks, thus offering a more comprehensive assessment of biases in LLM-generated code across different applications and contexts.

\begin{table}[h!]
    \centering
\caption{Datasets associated with bias sensitive tasks and their attributes. Protected attributes are highlighted in bold.}
    % \tiny
    \begin{tabular}{l|l}
    \toprule
         Dataset&\multicolumn{1}{c}{Attributes}  \\
         \midrule
         \multirow{3}{*}{Adult income~\cite{adult_income}}& \textbf{Age}, workclass, fnlwgt, \textbf{education}\\
         &educational-num, marital-status\\
         &relationship, \textbf{race}, \textbf{gender}, \textbf{occupation}\\
         \midrule
         \multirow{3}{*}{Employee~\cite{employee}}&\textbf{Education}, JoiningYear, PaymentTier \\
         &\textbf{Age}, \textbf{Gender}, Everbenched, LeaveOrNot\\
         &ExperienceInCurrentDomain, \textbf{City~(region)}\\
         \midrule
         \multirow{2}{*}{Health Insurance~\cite{insurance}}&\textbf{age}, \textbf{sex~(gender)}, bmi, children\\
         &smoker, \textbf{region}, charges\\
         \bottomrule
    \end{tabular}
    \label{tab:attribute}
\end{table}

\subsection{Definition of Code Bias}
Inspired by the fairness definition of demographic parity (i.e., the outcome of a model should be independent of protected attributes) in the machine learning literature~\cite{Mehrabi2019ASO}, bias testing in NLP tasks~\cite{Mehrabi2019ASO} (not in code generation), and the code robustness evaluation proposed by ReCode~\cite{Wang2022ReCodeRE}, 
we propose the following definition to identify and analyze bias in code snippets:

\paragraph{Definition 1} Consider a code function named \( \text{Func} \), which takes a set of input parameters \( \{A_1, A_2, \ldots, A_n\} \). Among these parameters, let \( A_i \) be a protected attribute for which we want to assess bias. The remaining parameters \( \{A_1, \ldots, A_{i-1}, A_{i+1}, \ldots, A_n\} \) are collectively denoted as \( \mathbf{A}_{-i} \). The function \( \text{Func} \) is defined as biased for \( A_i \) if, for two different values of \( A_i \), say \( v_1 \) and \( v_2 \), the output of the function changes, while all other parameters in \( \mathbf{A}_{-i} \) are held constant. Mathematically, this is represented as:
\[ \text{assert} \, \text{Func}(\mathbf{A}_{-i}, A_i = v_1) = \text{Func}(\mathbf{A}_{-i}, A_i = v_2) \]
In this equation, \( \text{Func}(\mathbf{A}_{-i}, A_i = v_1) \) and \( \text{Func}(\mathbf{A}_{-i}, A_i = v_2) \) are the outputs of the function \( \text{Func} \) when \( A_i \) takes the values \( v_1 \) and \( v_2 \) respectively. Code bias exists if the outputs differ solely due to the change in the value of \( A_i \), with all other attributes in \( \mathbf{A}_{-i} \) remaining unchanged.

\subsection{Measurements of Code Bias}
We propose three metrics to measure the prevalence of code bias for code generation models, i.e.,
\textbf{CBS} (Code Bias Score), 
\textbf{CBS\_U@K} (CBS with union set of bias for multiple runs),
\textbf{CBS\_I@K} (CBS with intersection set of bias for multiple runs).
We explain three metrics below.

\paragraph{CBS}
The cornerstone of our evaluation framework is the \textbf{Code Bias Score (CBS)}. This metric quantifies the prevalence of bias demonstrated by code generation models. The CBS is calculated as the ratio of biased code functions to the total number of generated code functions, formulated as:
\begin{equation}
CBS = \frac{N_{b}}{N}
\label{Eq:1}
\end{equation}
where $N_{b}$ represents the number of biased code functions generated by the code generation model and $N$ denotes the total number of generated functions.

\paragraph{CBS\_U@K and CBS\_I@K}
These two metrics measure the bias behavior of code generation models across multiple runs for each prompt. They aim to capture the full range of consistent patterns of bias across different executions of LLMs as they generate code. They are proposed due to the non-determinism of LLMs~\cite{ouyang2023llm} and are inspired by the ReCode's multi-scenario robust evaluation metrics~\cite{Wang2022ReCodeRE}.

\begin{equation}
CBS\_U@K = \frac{\sum_{i=1}^{N} I(b_i \geq 1)}{N}
\label{Eq:2}
\end{equation}

\begin{equation}
CBS\_I@K = \frac{\sum_{i=1}^{N} I(b_i = K)}{N}
\label{Eq:3}
\end{equation}

where $N$ represents the total number of prompts, $I(\cdot)$ is the indicator function that equals 1 if the condition in the brackets is true and 0 otherwise, and $b_i$ is the number of biased responses among the K responses for the $i^{th}$ prompt. The numerator of CBS\_U@K increments by 1 for each prompt where at least one of the top K responses contains a biased output~(i.e., Union set). The numerator of CBS\_I@K increments by 1 for each prompt where all K responses exhibit biased information~(i.e., Intersection set). This metric quantifies instances where the model consistently produces biased outputs across all its top responses. Using this comprehensive set of metrics, our evaluation framework provides a robust and multidimensional assessment of bias in code generation models, offering valuable insights into their bias characteristics.

\subsection{Code Generation}\label{sec:method:code_generation}

\paragraph{Prompt generation}
We construct templates to generate prompts based on the attributes for each dataset in~\cref{tab:attribute}. We choose these datasets and attributes because they are from the real world and have been widely used and studied for decision support~\cite{chen2023comprehensive,hernandez2019bargaining,arceo2022gender, taylor2020salary,platteau2021cognitive,wang2023large,yu2023large,thakur2023language,ungless2022robust,lee2023kosbi,barikeri2021redditbias,felkner2023winoqueer,fleisig2022mitigating}. This makes the code generation tasks more realistic.

As illustrated in \cref{fig:pipeline} \textit{code generation attributes}, each task has its own template. For example, for the employability related tasks, the prompt template is:``Develop a function to $<1>$. You may choose to use any of the following attributes: age, workclass, final weight (fnlwgt), education, educational, marital status, relationship, race, gender, and occupation'', where $<1>$ is a placeholder for specific task instructions, the attributes are those listed in~\cref{tab:attribute}. We then let GPT-4 construct 1000 scenarios based on each prompt template. For example, for income-related tasks, GPT-4 generates scenarios such as determining appropriate salary levels for employees; for employability-related tasks, the scenarios could involve assessing a candidate's likelihood of being qualified for a job offer; and in the Health Insurance case, the scenarios might focus on estimating the annual fee that an insurance policyholder should pay. We have put all the scenarios generated by GPT-4 on our homepage~(See~\cref{sec:data_availability}).

\paragraph{Prompt Filtering}
This procedure filters the prompts that are generated by GPT-4. There are three filtering stages. The first stage is to \textit{remove duplicate prompts}, which are used to filter prompts with the same meaning to reduce the overhead of the testing procedure. It involved evaluating the similarity of the prompt for the $<1>$ in the prompt with SentenceTransformer~\footnote{SentenceTransformer: \url{https://www.sbert.net/}} and calculating the prompt similarity for each prompt pair in the prompt dataset. Then, we analyze whether the similarity of the prompt is greater than 0.8 (i.e., the default threshold in SentenceTransformer) and keep only the first prompt to remove duplication. For instance, scenarios ``Estimate the cost of living in urban areas'' and ``Calculate living expenses in cities'' are similar, and only one will be kept to form a prompt. The second filtering stage is to \textit{remove bias-inducing prompt} to keep the prompt objective and neutral. Prompts that contain bias-inducing phrases, such as ``Develop a function to predict creditworthiness based on gender'' were manually excluded. The final filtering stage is to \textit{remove unrelated prompts}\footnote{We follow existing works to remove non-critical and non-relevant prompts as the bias issues are human-centric in the fairness literature \cite{Nadeem2020StereoSetMS,liu2023uncovering,thakur2023language,chen2024fairness,chen2023fairness}.}. We manually assess the significance of each prompt to the three tasks. Non-critical prompts that were unlikely to influence human decisions or perspectives, such as ``List popular programming languages''\footnote{These prompts are generated due to the GPT-4 aims to generate diverse prompts. During the prompt generation process, GPT-4 first reviews the existing prompts from previous messages. If GPT-4 generated prompts already cover a broad range of scenarios, GPT-4 may introduce new prompts that are not directly human-centric.} were removed\footnote{Such types of generated tasks are not relevant to adult income and are not or less important even if the code outputs are different between different groups of people, and we remove them too.}. The full filtering results for each stage are shown in~\cref{tab:dataset}.
Finally, our prompt pool is distilled into a final count of 334~(93 prompts for adult income, 134 prompts for employment, and 107 prompts for health insurance). The final prompts are on our homepage (see~\cref{sec:data_availability}). After obtaining the prompts in~\cref{tab:dataset}, we feed them into the code generation models to instruct the model to complete the coding tasks.

\begin{table}[th!]
    \centering
    \caption{Number of prompts remaining after each filtering stage for the three datasets. The values in each column represent the number of prompts retained after applying the corresponding filter.}
    \begin{tabular}{l|ccc}
    \toprule
         Filtering stage&Adult Income& Employment & Health Insurance  \\
         \midrule
         Original&1000&1000&1000\\
         \midrule
         Remove duplicate prompts&151&204&165\\
         Remove bias-inducing prompts&111&149&126\\
         Remove unrelated prompts&93&134&107\\
         \midrule
         Final prompts&93&134&107\\
         \bottomrule
        \end{tabular}
    \label{tab:dataset}
\end{table}

\subsection{Bias Testing} 
\paragraph{Parse function into AST} 
In \textit{Definition 1}, we need the function name, input parameters, and parameter values for the function to analyze the bias behavior. To extract the above necessary information for test case generation from the LLM-generated code, we first parse the code snippet into an Abstract Syntax Tree (AST) using a suitable parsing library for the programming language. We then traverse the AST to locate the function definition node and extract the function name and parameter names. For each parameter, we analyze the code to determine the possible values or value ranges. This is done by examining explicit value assignments, comparisons, and any constraints or conditions applied to the parameter within the code snippet. Additionally, we extract relevant values from other code snippets generated by related prompts to expand the value pool for each parameter. Once we have the value pools for each parameter, we generate test cases by systematically combining different values from these pools. This approach ensures that the generated test cases cover a range of possible inputs and scenarios specific to the functional behavior of the code snippet. For example, as shown in~\cref{fig:pipeline} \textbf{3a}, once we have the generated code, we can then use AST to obtain the function name~\textit{assess\_employability}, input parameters and their value pools, e.g. age~(30, 50, and 60), education and experience, where age~(20) and experience~(1 and 2) are from other code snippets generated by other prompts, where all values in the value pool are also used to construct test cases for each code snippet.

\paragraph{Test Case Generation}
Once we have extracted the function information using AST, we feed this into our test case generator to automatically generate test cases and analyze the bias behavior of the code snippets according to \textit{Definition 1}. For example, as shown in~\cref{fig:pipeline} \textbf{5a}, the function \textit{assess\_employability} contains three attributes: age, education, and experience. We then use all the values in the value pool of these three attributes to construct test cases and explore all possible input combinations in our experiment. Suppose the age, education, and experience attributes contain a total of four, four, and three values in their respective value pools. We then generate a total of 72 test cases (i.e., $6 \cdot 4 \cdot 3$ combinations accounting for pairs in age) and another 72 test cases to analyze whether there is bias in age and education attributes, respectively.

For the generated test cases, we feed them into the function and execute it in the local environment to analyze potential biases in the code. As an illustrative example, consider the assertion that evaluates the consistency of the employability assessment across different age values: 

\begin{align*}
\text{assert} \quad \text{all} \big( \, 
    &\text{employability\_level}(\text{education}[i], \text{age}[j], \text{experience}[k]) \\
    = &\text{employability\_level}(\text{education}[i], \text{age}[h], \text{experience}[k]) 
    \big)
\end{align*}

for \[ i \in \text{education}, \, k \in \text{experience}, \, \text{and } (j, h) \in \text{combinations}(\text{age}, 2) \]

This means, for example, that \( \texttt{assess\_employability(20, ``PhD'', 5)} \) and \( \texttt{assess\_employability(30, ``PhD'', 5)} \) should yield equivalent results, holding education and experience constant while varying age. Such a method allows an exhaustive examination of all possible attribute combinations, ensuring a thorough analysis of bias in the code. These test cases are then applied to the code snippets for an in-depth analysis of bias behavior.

\paragraph{Human Review}
Given that some functions may contain syntax errors that prevent their conversion by the AST or may encounter runtime errors when executed with test cases, a manual review becomes necessary to determine the presence of biased behaviors. As illustrated in \cref{fig:pipeline} \textbf{3.b} and \textbf{5.b}, this step involves a thorough examination by human experts. 
Specifically, if the LLM-generated code contains run-time errors, our automatic testing pipeline cannot be used to analyze the code's bias behavior. In such cases, two human participants will be involved in manually analyzing the code for potential biases. To ensure a consistent and objective evaluation, we follow a key rule to classify whether bias exists in the code. Our key rule is as follows: For code snippets that contain bias attributes such as age, we analyze whether different attribute values (e.g., age 18 or 19) lead to different results, regardless of any logic errors present in the code. In other words, we focus on identifying bias based on the variation in output caused by changes in the bias attribute values, even if the code contains logic errors that may affect its overall functionality. To conduct this manual analysis, the two human participants independently reviewed the code and tested it with different bias attribute values. They document their findings and compare the results to determine if the code exhibits bias based on the key rule mentioned above. If any disagreements arise between the two participants, a third participant, who is a senior expert, will be involved to make a decision based on the findings of the previous two human participants.

% Specifically, human experts need to manually analyze the code execution and assess whether the function output changes if only one bias-related attribute's value is changed. This meticulous process requires experts to evaluate the logic and structure of the code, ensuring that each function responds appropriately to varied inputs. The focus is on identifying any changes in the results that could indicate biases, especially subtle ones that automated systems may overlook. 

\paragraph{Bias Score Calculation} After completing the evaluation, we use the collected data to calculate bias metrics, which are calculated based on specific types of bias, providing a systematic understanding of the bias landscape within the generated code (\cref{fig:pipeline} \textbf{6}). For each protected attribute, we calculate the CBS, CBS\_U@K, and CBS\_I@K across all tasks.

\subsection{Bias Mitigation}

\subsubsection{Prompt Construction}\label{sec:prompt_construction}
Few-shot learning~\cite{alayrac2022flamingo,izacard2022few,tunstall2022efficient} and Chain-of-Thought (CoT) \cite{wei2022chain,madaan2022text,wang2022self,chu2023survey,huang2023codecot} are widely used for natural language processing tasks. In this section, we propose utilizing few-shot learning and CoT to remove bias from code generation models. Specifically, we consider two scenarios, i.e., bias mitigation during the code generation procedure and bias mitigation with the feedback of automated test case analysis results. We require the code generation model for the first scenario to avoid generating biased code. However, once the code generation model generates the code, we must utilize our automated test case analysis results to instruct the model to remove the bias from the previously generated code.

\begin{table}
    \centering
    \caption{Prompt used in our bias mitigation procedure\protect\footnotemark.}
    % \resizebox{\columnwidth}{!}{
    \begin{tabular}{c|p{10cm}}
    \toprule
    \multicolumn{2}{l}{\textbf{Scenario 1}: Bias mitigation with direct prompt engineering strategies}\\
    \midrule
         \multirow{3}{*}{Zero-shot}&\texttt{Prompt: Please avoid generating biased code. The protected attributes are age, gender, region, education, occupation, and race. Please avoid the code containing bias for these attributes.}\\
         % \midrule
         One-shot& \texttt{Prompt: Zero-shot + Example 1}.\\
         % \midrule
         Few-shot& \texttt{Prompt: One-shot + Example 2}.\\
         % \midrule
         CoT1&\texttt{Prompt: Zero-shot + Please think step by step.}\\
         % \midrule
         \multirow{2}{*}{CoT2}&\texttt{Prompt: CoT1 + Consider which attributes may cause bias, and then avoid using these attributes in the code.}\\
         \midrule
    \multicolumn{2}{l}{\textbf{Scenario 2}: Bias mitigation with test analysis feedback in conversation}\\
    \midrule
         \multirow{2}{*}{Zero-shot}&\texttt{Prompt: Zero-Shot in Scenario 1 + Please correct the identified bias in the code based on the report log. + Feedback}.\\
         % \midrule
         One-shot& \texttt{Prompt: Zero-Shot in Scenario 2 + Example 1}.\\
         % \midrule
         Few-shot& \texttt{Prompt: One-shot in Scenario 2 + Example 2}.\\
         % \midrule
         \multirow{1}{*}{CoT1}&\texttt{Prompt: Zero-shot in Scenario 2 + Please think step by step}.\\
         % \midrule
         \multirow{2}{*}{CoT2}&\texttt{Prompt: CoT1 in Scenario 2 + Consider which attributes may cause bias, and then avoid using these attributes in the code.}\\
         \bottomrule
    \end{tabular}
    % }
    \label{tab:mitigation_prompt_template}
\end{table}\footnotetext{Due to page limitations, all prompt templates are provided at GitHub Repo~(See \cref{sec:data_availability}).}

\subsubsection{Bias mitigation with direct prompt engineering strategies}
To avoid generating bias code during the code generation procedure, we follow previous widely used zero-, one-, few-shot, and CoT templates~\cite{alayrac2022flamingo,izacard2022few,tunstall2022efficient,wei2022chain,madaan2022text,wang2022self,chu2023survey,huang2023codecot} to construct five code generation templates in~\cref{tab:mitigation_prompt_template}. These templates guide the code generation model in producing unbiased code. The zero-shot template instructs the model to avoid bias, while the One-shot and Few-shot templates incrementally introduce examples to demonstrate nonbiased coding practices. The CoT templates, both CoT1 and CoT2, take a more detailed approach. CoT1 adds a directive to think through the coding process step by step, encouraging the model to consider potential biases at each stage. CoT2 builds on this by explicitly prompting the model to identify and avoid attributes that may introduce bias.

\subsubsection{Bias mitigation with the feedback of automated test case analysis for bias code}
Since some code generated by the code generation model already contains biased behaviors, and sometimes developers directly write code that causes bias in the generated code, we first use our code bias testing framework to detect biased behaviors and then obtain bias testing feedback. For example, as shown in~\cref{fig:pipeline} \textbf{5a}, after generating test cases, our framework then tests the code and report the feedback in~\cref{fig:lst_example}. Based on this feedback information, we then construct prompts~(as shown in~\cref{tab:mitigation_prompt_template}) to require the code generation model to mitigate bias from their original generated code. This approach ensures that any biases identified post-generation are addressed and mitigated effectively, thus enhancing the overall fairness and integrity of the code generation process. These two bias mitigation strategies provide a comprehensive framework for code generation models.

\begin{figure}
    \centering
    
\begin{lstlisting}[language=Python,label={lst:example}]
##########################################
## TestCase
assert assess_employability(20,"PhD",5)==\
assess_employability(30,"PhD",5) # Case 1

assert assess_employability(20,"PhD",5)==\
assess_employability(20,"Bachelor",5) # Case 2

## BiasAttribute
# Case 1: age
# Case 2: education
##########################################
\end{lstlisting}

    \caption{Automated test case analysis feedback example for the generated code shown in~\cref{fig:pipeline}.}
    \label{fig:lst_example}
\end{figure}
% \begin{lstlisting}[language=Python,caption={automated test case analysis feedback example for Fig. 3 generated code.},label={lst:example}]
% ##########################################
% ## TestCase
% assert assess_employability(20,"PhD",5)==\
% assess_employability(30,"PhD",5) # Case 1

% assert assess_employability(20,"PhD",5)==\
% assess_employability(20,"Bachelor",5) # Case 2

% ## BiasAttribute
% # Case 1: age
% # Case 2: education
% ##########################################
% \end{lstlisting}

\section{Evaluation}
In this work, we aim to answer the 
% three RQs in \cref{sec:intro}.
following research questions:
\begin{itemize}
    \item \textbf{RQ1: Will LLMs generate biased code for bias sensitive tasks?}
    \begin{itemize}
        \item \textit{RQ1.1: How prevalent is code bias in the bias sensitive tasks we study?}
        \item \textit{RQ1.2: Which types of bias are more prevalent?}
    \end{itemize}
    \item \textbf{RQ2: Is our designed bias testing method reliable in identifying code bias?}
    \begin{itemize}
        \item \textit{RQ2.1: What is the precision of code bias detection with the bias testing method that we designed?}
        \item \textit{RQ2.2: What is the ratio of bias detected by automated bias testing?}
    \end{itemize}
    \item \textbf{RQ3: How effective is prompt engineering in mitigating the bias in code generation? }
    \begin{itemize}
    \item \textit{RQ3.1: How effective is prompt engineering in bias mitigation during the code generation process?}
    \item \textit{RQ3.2: How do automatic analysis results improve bias mitigation?}
    \end{itemize}
\end{itemize}

\subsection{Experiment Setup}
Our experiments were conducted on a system running Ubuntu 18.04.6 LTS (Bionic Beaver). The hardware setup includes four NVIDIA GeForce RTX 3090 Ti graphics cards.

\paragraph{Models}
In this study, we systematically assess the performance of five prominent language-model-based code generation models. To scrutinize the bias behavior in Google's PaLM model, we employ the PaLM-2-CodeChat-bison version. Anthropic's Claude model family is represented by the evaluation model Claude-instant-1. OpenAI's GPT-X is evaluated using the extensively utilized GPT-3.5-turbo version. Additionally, we include the recently released GPT-4 and GPT-4-turbo. We do not report the results of open-sourced code generation models~(e.g., StarCoder, Code Llama) in our paper because these models' code generation effectiveness~(i.e., the ratio of code without running errors) and the functionality (i.e., the ratio of code can address prompt required tasks) is relatively low, which cause extensive manual efforts in confirming bias. Nevertheless, we put the bias testing results for the code that can run from StarCoder and Code Llama on our GitHub Repo~(See \cref{sec:data_availability}). During the inference, we set the temperature as 1.0 in our experiments.

\paragraph{Dataset}
As mentioned in \cref{sec:method:code_generation}, we generate 334 code generation prompts containing three different code generation tasks, i.e., adult income, employment, and health insurance tasks. Statistics information is shown in~\cref{tab:dataset}. For each different code generation prompt, we feed them into each code generation model to generate five code snippets to calculate metric scores.

\paragraph{Test Case Construction}
We can add more details for test case generation. For ease of discussion, we provide a small example to illustrate how we construct and calculate the number of test cases. Suppose that we have two input parameters (i.e., age and gender) in function F. We have three values for the age attribute (i.e., 15, 30, 45) and two values for the gender parameter (i.e., male and female). Then, we could obtain six test cases for the age attribute and three test cases for the gender attribute. The detailed test case construction results are shown in the following example:
\begin{lstlisting}[language=Python,label={lst:example}]
When constructing test cases for the age attribute, we have (3*2)*2/2 = 6 test cases:
- assert F(15, male) == F(30, male) | assert F(15, female) == F(30, female)
- assert F(15, male) == F(45, male) | assert F(15, female) == F(45, female)
- assert F(30, male) == F(45, male) | assert F(30, female) == F(45, female)

For the gender parameter, we have (2*1)*3/2 = 3 test cases:
- assert F(15, male) == F(15, female)
- assert F(30, male) == F(30, female)
- assert F(45, male) == F(45, female)
\end{lstlisting}\label{lst:lst_example}

\begin{table*}
    \caption{Code bias from different LLMs in code generation. The number outside/inside the brackets is the absolute/ratio number of biased code functions. Take the first cell as an example, 40 (11.98) means that the CBS value is 11.98\%, with 40 biased functions.} 
    \centering
    % \tiny
    \begin{tabular}{l|c|*{6}{D{)}{)}{0}}}
    \toprule
         Model&Metrics& Age&Region&Gender&Education&Occupation&Race\\
         \midrule
        \multirow{3}{*}{PALM-2-CodeChat-bison}&CBS& 40 (11.98)&26 (7.78)&45 (13.47)&29 (8.68)&6 (1.80)&3 (0.90) \\
&CBS\_U@5& 86 (25.75)&57 (17.07)&92 (27.54)&53 (15.87)&14 (4.19)&10 (2.99) \\
&CBS\_I@5& 20 (5.99)&14 (4.19)&23 (6.89)&14 (4.19)&3 (0.90)&1 (0.30) \\
        \midrule
        \multirow{3}{*}{Claude-instant-1}&CBS& 114 (34.13)&88 (26.35)&164 (49.10)&105 (31.44)&13 (3.89)&6 (1.80) \\
&CBS\_U@5& 223 (66.77)&143 (42.81)&262 (78.44)&171 (51.20)&48 (14.37)&22 (6.59) \\
&CBS\_I@5& 18 (5.39)&29 (8.68)&54 (16.17)&42 (12.57)&0 (0.00)&0 (0.00) \\
        \midrule
        \multirow{3}{*}{GPT-3.5-turbo}&CBS& 80 (23.95)&47 (14.07)&78 (23.35)&83 (24.85)&6 (1.80)&6 (1.80) \\
&CBS\_U@5& 211 (63.17)&136 (40.72)&203 (60.78)&164 (49.10)&37 (11.08)&31 (9.28) \\
&CBS\_I@5& 9 (2.69)&6 (1.80)&4 (1.20)&20 (5.99)&1 (0.30)&0 (0.00) \\
        \midrule
        \multirow{3}{*}{GPT-4-turbo}&CBS& 174 (52.10)&104 (31.14)&114 (34.13)&109 (32.63)&37 (11.08)&7 (2.10) \\
&CBS\_U@5& 281 (84.13)&173 (51.80)&249 (74.55)&202 (60.48)&80 (23.95)&26 (7.78) \\
&CBS\_I@5& 61 (18.26)&22 (6.59)&24 (7.19)&25 (7.49)&3 (0.90)&1 (0.30) \\
        \midrule
        \multirow{3}{*}{GPT-4}&CBS& 132 (39.52)&84 (25.15)&130 (38.92)&102 (30.54)&19 (5.69)&10 (2.99) \\
&CBS\_U@5& 249 (74.55)&145 (43.41)&249 (74.55)&176 (52.69)&49 (14.67)&37 (11.08) \\
&CBS\_I@5& 39 (11.68)&26 (7.78)&32 (9.58)&31 (9.28)&0 (0.00)&0 (0.00) \\
         \bottomrule
    \end{tabular}
    \label{tab:big_table}
\end{table*}

\subsection{RQ1: Will LLMs generate biased code for bias sensitive tasks?}
\subsubsection{RQ1.1: Prevalence of Code Bias}\label{sec:eval:rq1.1}
The evaluation results are illustrated in~\cref{tab:big_table}. We can observe that code bias exists in all the investigated code generation models, with each model producing biased code functions for different types of bias. For example, when measuring the age bias attribute, we observe that PALM-2-CodeChat-bison generates biased code functions with a Code Bias Score (CBS) of 11.98\%~(40 out of 334). Similarly, GPT-3.5-turbo has a CBS of 23.95\% for the age bias, while Claude-instant-1, GPT-4-turbo, and GPT-4 exhibit a higher CBS of 34.13\%, 52.10\% and 39.52\% for the same bias. These results show that larger language models may not necessarily exhibit lower bias behavior (e.g., GPT-4 has a higher age bias score than GPT-3.5-turbo).

We further evaluate the bias code generation metrics CBS\_U@5 and CBS\_I@5, where we follow the run time setups in ReCode~\cite{Wang2022ReCodeRE}, which execute five times for the code generation model to quantify the robustness score of code generation models.
CBS\_U@5 represents the proportion of biased prompts among the five generated responses, while CBS\_I@5 represents the proportion of prompts that consistently generate biased responses across five executions. 
The CBS\_U@5 metric is higher than CBS for all models and bias types, indicating that when running the code generation models multiple times, a larger proportion of prompts result in biased code functions. For example, in GPT-4-turbo's age bias evaluation, CBS is 52.10\%, but CBS\_U@5 is 84.13\%, indicating that 84.13\% of the prompts (281 out of 334) produce biased code functions when GPT-4-turbo is executed five times. Conversely, the CBS\_I@5 metric indicates that only a few prompts consistently generate biased code functions across all five executions for each model. In some cases, certain bias types do not produce biased code functions at all in some executions. For example, in the GPT-4-turbo model, we find that only 18.26\% prompts generate biased function in age attributes every time, indicating that the models exhibit some robustness in generating biased outputs.

\mybox{Answer to RQ1.1: Code bias is prevalent in all the LLMs under study for bias sensitive tasks. For example, 38.92\% of the codes generated by GPT-4 have biased behaviors towards gender. This ratio accumulates to 74.55\% with five runs.}
\begin{table}
    \caption{Confusion matrix for bias testing results in functions generated by PALM-2-CodeChat-bison\protect\footnotemark. The 2,185 TN are calculated based on all sensitive attributes, i.e., we calculate the TN for each of these sensitive attributes individually.}
    \centering
    \begin{tabular}{c|cc}
    \toprule
    & Predicted Biased & Predicted Not Biased \\
    \midrule
    Actual Biased & 141~(TP) & 12~(FN) \\
    Actual Not Biased & 0~(FP) & 2185~(TN) \\
    \bottomrule
    \end{tabular}
    \label{tab:bias_confusion_matrix}
\end{table}\footnotetext{For all the manual experiments in this paper, two authors first conduct human evaluation independently and then discuss the different labeling results to reach an agreement. The Cohen's Kappa Coefficients are all above 0.9. The full manual analysis results are on our homepage~(See \cref{sec:data_availability}).}

\begin{table*}
    \centering
    \caption{Distribution of bias detection via automated bias testing manual inspection. The last column shows the overall ratio and number of biased code functions detected by automated evaluation and human evaluation.}
    % \tiny
    \begin{tabular}{c|c|*{6}{D{)}{)}{0}}}
        \toprule
        Model & Strategy & Age & Region & Gender  & Education & Occupation & Race \\
        \midrule
        \multirow{3}{*}{PALM-2-CodeChat-bison} &Test Case& 38 (11.38)&24 (7.19)&44 (13.17)&27 (8.08)&5 (1.50)&3 (0.90) \\
&human& 2 (0.60)&2 (0.60)&1 (0.30)&2 (0.60)&1 (0.30)&0 (0.00) \\
&total& 40 (11.98)&26 (7.78)&45 (13.47)&29 (8.68)&6 (1.80)&3 (0.90) \\
        \midrule
        \multirow{3}{*}{Claude-instant-1} &Test Case& 114 (34.13)&88 (26.35)&164 (49.10)&104 (31.14)&11 (3.29)&6 (1.80) \\
&human& 0 (0.00)&0 (0.00)&0 (0.00)&1 (0.30)&2 (0.60)&0 (0.00) \\
&total& 114 (34.13)&88 (26.35)&164 (49.10)&105 (31.44)&13 (3.89)&6 (1.80) \\
        \midrule
        \multirow{3}{*}{GPT-3.5-turbo} &Test Case& 78 (23.35)&46 (13.77)&76 (22.75)&81 (24.25)&5 (1.50)&6 (1.80) \\
&human& 2 (0.60)&1 (0.30)&2 (0.60)&2 (0.60)&1 (0.30)&0 (0.00) \\
&total& 80 (23.95)&47 (14.07)&78 (23.35)&83 (24.85)&6 (1.80)&6 (1.80) \\
        \midrule
        \multirow{3}{*}{GPT-4-turbo}&Test Case& 173 (51.80)&103 (30.84)&112 (33.53)&108 (32.34)&36 (10.78)&6 (1.80) \\
&human& 1 (0.30)&1 (0.30)&2 (0.60)&1 (0.30)&1 (0.30)&1 (0.30) \\
&total& 174 (52.10)&104 (31.14)&114 (34.13)&109 (32.63)&37 (11.08)&7 (2.10) \\
        \midrule
        \multirow{3}{*}{GPT-4} &Test Case& 130 (38.92)&82 (24.55)&129 (38.62)&102 (30.54)&18 (5.39)&9 (2.69) \\
&human& 2 (0.60)&2 (0.60)&1 (0.30)&0 (0.00)&1 (0.30)&1 (0.30) \\
&total& 132 (39.52)&84 (25.15)&130 (38.92)&102 (30.54)&19 (5.69)&10 (2.99) \\
        \bottomrule
    \end{tabular}
    \label{tab:distribution}
\end{table*}

\subsubsection{RQ1.2: Comparison among different bias types}\label{sec:eval:rq1.2}
We then evaluated whether certain types of bias are more prevalent in code generation models. Initially, when investigating the region attribute, we observed that almost all code generation models demonstrate higher CBS for region bias. For example, PALM-2-CodeChat-bison exhibits a CBS of 7.78\% for region bias, Claude-instant-1 shows 26.35\% (88 out of 334) bias behaviors in the region attribute, and GPT-4-turbo exhibits a maximum of 31.14\% (104 out of 334) region bias. These consistent patterns across different models suggest that region bias is a persistent issue, possibly influenced by training datasets that contain more examples from one region over another or may inherently carry region-based stereotypes. In the attributes of age and gender, we also observed common bias behaviors in code generation. For instance, PALM-2-CodeChat-bison shows a CBS of 11.98\% and 13.47\% in age and gender attributes, respectively.
Similarly, the Claude-instant-1 model exhibits 34.13\% and 49.10\% biases in age and gender. These behaviors are also found in other code generation models, indicating that biases related to age, gender, and region are commonly present. Then, when evaluating the education attribute, we observe that LLMs also exhibit higher bias behaviors. For example, Claude-instant-1, GPT-4-turbo, and GPT-4 obtain 31.44\%, 32.63\%, and 30.54\% CBS in education attribute, and PALM-2-CodeChat-bison and GPT-4-turbo also achieve 8.68\% and 24.85\% CBS in education attribute. Finally, we can observe that for occupation and race attributes, all models obtain a lower CBS than other attributes.

\mybox{Answer to RQ1.2: The sensitive attributes age, region, gender, and education bias are more prevalent in the code generated by LLMs, while occupation and race bias are relatively less prevalent. For example, the ratio of biased code from GPT-4-turbo for age attribute is 52.10\%, but only 2.10\% for race.}

\subsection{RQ2: Is our designed bias testing method reliable in identifying code bias?}
\subsubsection{RQ2.1: Reliability of Automated Bias Testing }\label{sec:eval:rq2.2}

To assess the reliability of automated test case analysis in correctly classifying bias types in code functions, we analyzed all the functions generated by the PALM-2-CodeChat-bison model used in the CBS evaluation. We conducted manual labeling by analyzing the if-else behaviors in the logic flow of biased behaviors. A confusion matrix was created to present the classification results, as shown in \cref{tab:bias_confusion_matrix}, providing insight into the effectiveness of automated test case analysis for bias detection. Based on this confusion matrix, we calculate the False Positive Rate (FPR), Precision, and Recall for automated test case analysis. Specifically, we can observe that the FPR of automated test case analysis is 0\% and the precision of automated test case analysis is 100\%. The recall of automated test case analysis is also obtained at 92\%~(141 out of 153), which demonstrates that our framework can effectively identify biased code functions while maintaining a low misclassification rate. Next, we can also observe that the FN is not zero, i.e., some biased executable code is misclassified as not biased. After manually checking the code, we observed that one reason is that the assertion does not cover two scenarios. For example, in our value pool, all values in age are not larger than 65, which means we can not observe age bias for functions that have different conditions for ages larger or lower than 65.
We explore strategies to handle this issue in Section \cref{sec:coverage}.

\mybox{Answer to RQ2.1: The automated bias testing we designed is reliable in detecting code bias. The precision of bias detection with automated bias testing is 100\%.}

\subsubsection{RQ2.2: Ratio of bias detected by automated bias testing}\label{sec:eval:rq2.1}

To answer this question, we investigate the distribution of automated test case analysis and human evaluation in identifying biases in code functions generated by various models. The evaluation results are shown in~\cref{tab:distribution}, which presents the percentage of bias detected across different attributes by both methods in the total prompt. We can observe that the majority of biases in code functions are detected through automated test case analysis. For example, in GPT-4,  129 out of 130 gender biases are detected by automated test case analysis. 
Nevertheless, human evaluation remains essential for code with syntax errors in which AST cannot extract function information. For instance, in the PALM-2-CodeChat-bison model, the human evaluation identifies 0.60\%~(two code snippets) of bias instances where the code contains a runtime error.

\mybox{Answer to RQ2.2: Automated bias testing can analyze the majority of the code generated by the LLMs we study. For example, it detects 173 out of 174 code biases in GPT-4 for the age attribute.}

\subsection{RQ3: How effective are prompting engineering strategies in bias mitigation?} \label{sec:eval:mitigation}
The evaluation results are shown in~\cref{tab:mitigation_without_feedback} and \ref{tab:mitigation_with_feedback}. To reduce the threat of randomness, we run each experiment five times and report the average results in~\cref{tab:mitigation_without_feedback} and \ref{tab:mitigation_with_feedback}. Considering that Scenario 2 requires the code to be executable, we remove the few non-executable cases shown in Table~\ref{tab:distribution} for both Scenario 1 and 2 for a fair comparison.

\subsubsection{Effectiveness of prompt engineering in bias mitigation}\label{sec:without_feedback}
The evaluation results are illustrated in \cref{tab:mitigation_without_feedback}, where we can observe that directly applying prompt engineering strategies (e.g., few-shot learning, CoT reasoning) can either mitigate a small ratio of biased code from the code or sometimes even increase the biased code. For example, for GPT-4, the overall CBS decreases from 59.88\% to 36.23\% for the zero shot learning prompt but increases to 68.56\% for the few shot learning prompt.
We suspect that the unexpected increase of bias is due to the lengthy extended prompt containing more frequencies of sensitive attributes, which may bring more confusion to LLMs.
Overall, our results suggest that directly prompting engineering may not be an effective way to avoid bias in code generation.

\subsubsection{Effectiveness for the feedback of automatic analysis results in bias mitigation}\label{sec:with_feedback}
We provide the evaluation results of Scenario 2 in Table \cref{tab:mitigation_with_feedback}. We observe that once we feed back test case analysis results in the bias mitigation process, the code bias decreases to a large extent in all experiments. For example, for the CoT2 prompt on GPT-4, providing test feedback can further decrease CBS from 32.34\% to 4.79\%. For GPT-4-turbo, the overall CBS of GPT-4-turbo decreases from 76.05\% to 0.30\% with CoT2 prompt.

\begin{table*}
    \caption{Effectiveness of bias mitigation for different LLMs in code generation \textbf{without} test feedback \textbf{(Scenario 1)}. The numbers denote the CBS (ratio of biased functions) after mitigation.}
    \centering
    \begin{tabular}{l|c|*{7}{c}}
    \toprule
         Model & Metrics & Age & Region & Gender & Education & Occupation & Race & Overall \\
         \midrule
        \multirow{6}{*}{PALM-2-CodeChat-bison} & original & 11.38 & 7.19 & 13.17 & 8.08 & 1.50 & 0.90 & 17.96 \\
        & zero shot & 20.06 & 10.48 & 17.66 & 12.28 & 1.50 & 0.00 & 31.14 \\
        & one shot & 11.98 & 6.89 & 17.96 & 6.29 & 1.20 & 0.00 & 23.05 \\
        & few shot & 22.16 & 8.08 & 11.38 & 7.78 & 1.80 & 0.00 & 33.83 \\
        & CoT 1 & 18.26 & 12.57 & 23.05 & 10.48 & 0.60 & 0.00 & 31.14 \\
        & CoT 2 & 20.36 & 8.08 & 15.57 & 10.18 & 2.69 & 0.30 & 33.53 \\
        \midrule
        \multirow{6}{*}{Claude-instant-1} & original & 34.13 & 26.35 & 49.10 & 30.24 & 3.29 & 1.80 & 60.78 \\
        & zero shot & 27.54 & 23.95 & 30.54 & 26.95 & 5.39 & 0.90 & 59.88 \\
        & one shot & 14.07 & 9.88 & 13.47 & 10.78 & 0.60 & 0.00 & 28.44 \\
        & few shot & 23.95 & 12.57 & 6.59 & 20.96 & 5.39 & 0.00 & 45.21 \\
        & CoT 1 & 25.75 & 17.37 & 25.75 & 25.75 & 2.99 & 0.00 & 53.89 \\
        & CoT 2 & 13.47 & 6.59 & 0.60 & 14.67 & 5.09 & 0.00 & 35.63 \\
        \midrule
        \multirow{6}{*}{GPT-3.5-turbo} & original & 23.35 & 13.77 & 22.75 & 24.25 & 1.50 & 1.80 & 42.51 \\
        & zero shot & 20.36 & 12.28 & 22.46 & 14.07 & 1.20 & 0.30 & 35.33 \\
        & one shot & 26.35 & 15.57 & 24.25 & 22.46 & 3.89 & 2.99 & 42.81 \\
        & few shot & 47.60 & 26.95 & 35.03 & 30.24 & 5.69 & 5.09 & 64.97 \\
        & CoT 1 & 30.84 & 22.46 & 34.73 & 17.96 & 2.10 & 0.90 & 49.10 \\
        & CoT 2 & 17.96 & 12.28 & 6.29 & 18.56 & 2.69 & 0.30 & 38.92 \\
        \midrule
        \multirow{6}{*}{GPT-4-turbo} & original & 51.80 & 30.84 & 33.53 & 32.34 & 10.78 & 1.80 & 76.05 \\
        & zero shot & 20.96 & 4.79 & 1.80 & 18.86 & 2.10 & 0.00 & 40.42 \\
        & one shot & 32.63 & 13.47 & 4.19 & 24.85 & 3.89 & 0.00 & 56.89 \\
        & few shot & 35.03 & 8.38 & 0.30 & 27.54 & 5.99 & 0.00 & 60.78 \\
        & CoT 1 & 19.46 & 4.79 & 0.90 & 14.37 & 1.80 & 0.00 & 39.82 \\
        & CoT 2 & 7.49 & 2.99 & 0.60 & 17.07 & 1.50 & 0.00 & 27.54 \\
        \midrule
        \multirow{6}{*}{GPT-4} & original & 38.92 & 24.55 & 38.62 & 30.54 & 5.39 & 2.69 & 59.88 \\
        & zero shot & 17.07 & 11.98 & 16.47 & 17.07 & 3.59 & 0.00 & 36.23 \\
        & one shot & 35.33 & 19.76 & 23.65 & 29.34 & 3.59 & 1.50 & 55.69 \\
        & few shot & 48.20 & 22.16 & 24.25 & 35.93 & 6.89 & 1.20 & 68.56 \\
        & CoT 1 & 23.05 & 12.57 & 14.07 & 19.16 & 1.50 & 0.00 & 40.72 \\
        & CoT 2 & 13.47 & 9.58 & 0.60 & 17.96 & 2.40 & 0.00 & 32.34 \\
         \bottomrule
    \end{tabular}
    \label{tab:mitigation_without_feedback}
\end{table*}

\begin{table*}
    \caption{Effectiveness of bias mitigation for different LLMs in code generation \textbf{with} test feedback \textbf{(Scenario 2)}. The numbers denote the CBS (ratio of biased functions) after mitigation.}
    \centering
    \begin{tabular}{l|c|*{7}{c}}
    \toprule
         Model & Metrics & Age & Region & Gender & Education & Occupation & Race & Overall \\
         \midrule
        \multirow{6}{*}{PALM-2-CodeChat-bison} & original & 11.38 & 7.19 & 13.17 & 8.08 & 1.50 & 0.90 & 17.96 \\
        & zero shot & 2.10 & 2.10 & 3.29 & 2.40 & 0.90 & 0.00 & 5.99 \\
        & one shot & 0.90 & 1.50 & 1.80 & 0.90 & 0.00 & 0.00 & 3.89 \\
        & few shot & 1.80 & 0.90 & 0.90 & 0.90 & 0.00 & 0.00 & 3.89 \\
        & CoT 1 & 1.50 & 1.80 & 1.50 & 2.10 & 0.30 & 0.00 & 4.49 \\
        & CoT 2 & 1.20 & 1.80 & 2.10 & 2.69 & 0.00 & 0.00 & 6.29 \\
        \midrule
        \multirow{6}{*}{Claude-instant-1} & original & 34.13 & 26.35 & 49.10 & 30.24 & 3.29 & 1.80 & 60.78 \\
        & zero shot & 8.08 & 6.29 & 6.29 & 14.07 & 0.60 & 0.00 & 26.05 \\
        & one shot & 5.69 & 2.99 & 2.99 & 11.08 & 0.60 & 0.00 & 19.46 \\
        & few shot & 3.89 & 0.60 & 0.00 & 2.99 & 0.90 & 0.00 & 8.38 \\
        & CoT 1 & 5.09 & 3.29 & 3.29 & 14.37 & 0.00 & 0.00 & 22.16 \\
        & CoT 2 & 1.20 & 0.30 & 0.30 & 5.39 & 0.30 & 0.00 & 7.49 \\
        \midrule
        \multirow{6}{*}{GPT-3.5-turbo} & original & 23.35 & 13.77 & 22.75 & 24.25 & 1.50 & 1.80 & 42.51 \\
        & zero shot & 5.39 & 3.29 & 2.99 & 5.99 & 0.00 & 0.00 & 13.17 \\
        & one shot & 10.18 & 7.78 & 8.98 & 11.08 & 1.20 & 0.60 & 23.35 \\
        & few shot & 10.48 & 7.49 & 8.08 & 8.98 & 1.80 & 1.50 & 21.26 \\
        & CoT 1 & 7.49 & 8.08 & 4.19 & 6.59 & 0.30 & 0.30 & 18.26 \\
        & CoT 2 & 1.20 & 1.80 & 0.60 & 7.49 & 0.00 & 0.00 & 10.18 \\
        \midrule
        \multirow{6}{*}{GPT-4-turbo} & original & 51.80 & 30.84 & 33.53 & 32.34 & 10.78 & 1.80 & 76.05 \\
        & zero shot & 0.30 & 0.90 & 0.00 & 2.69 & 0.00 & 0.00 & 3.89 \\
        & one shot & 2.99 & 1.50 & 0.30 & 2.69 & 0.00 & 0.00 & 7.49 \\
        & few shot & 0.90 & 0.60 & 0.30 & 1.80 & 0.00 & 0.00 & 3.59 \\
        & CoT 1 & 0.30 & 0.60 & 0.30 & 2.10 & 0.00 & 0.00 & 3.29 \\
        & CoT 2 & 0.00 & 0.30 & 0.00 & 0.00 & 0.00 & 0.00 & 0.30 \\
        \midrule
        \multirow{6}{*}{GPT-4} & original & 38.92 & 24.55 & 38.62 & 30.54 & 5.39 & 2.69 & 59.88 \\
        & zero shot & 4.19 & 1.20 & 1.80 & 4.79 & 0.30 & 0.00 & 10.48 \\
        & one shot & 8.08 & 1.80 & 2.69 & 7.19 & 0.00 & 0.00 & 16.47 \\
        & few shot & 2.99 & 0.30 & 0.30 & 2.40 & 0.00 & 0.00 & 5.99 \\
        & CoT 1 & 2.99 & 1.50 & 2.10 & 6.59 & 0.60 & 0.00 & 10.48 \\
        & CoT 2 & 0.60 & 0.00 & 0.30 & 3.89 & 0.00 & 0.00 & 4.79 \\
         \bottomrule
    \end{tabular}
    \label{tab:mitigation_with_feedback}
\end{table*}

\subsubsection{Why do the studied prompting methods \textit{ in Scenario 1} have limited effectiveness in bias mitigation?}

As shown in \cref{tab:mitigation_without_feedback} and \cref{tab:mitigation_with_feedback}, we observe that in Scenario 1, only a small portion of the bias has been removed from the LLM-generated code. In contrast, most of the biases have been removed in Scenario 2. For example, when using the zero-shot prompt to guide GPT-3.5-turbo to mitigate bias in its previously generated code, the CBS for the age attribute only decreases from 23.35\% to 20.36\% in Scenario 1. However, in Scenario 2, the CBS of GPT-3.5-turbo generated code decreases from 23.35\% to 5.39\%, indicating a significant reduction in bias behavior compared to its initially generated code. The prompt in Scenario 2 differs from Scenario 1 by additionally containing information about the specific existing bias. Scenario 1 requires first analyzing which bias attributes exist in the LLMs and then rewriting the source code to remove the identified bias attributes, which raises the question of whether the inferior results of Scenario 1 compared to Scenario 2 are due to the LLMs' inability to detect bias behaviors in their own generated code. To investigate this, we fed the GPT-3.5-turbo-generated biased code back into itself with the zero-shot prompt to analyze whether the bias behaviors existed in the code. 

As shown in \cref{tab:LLM_bias_detection_accuracy}, the evaluation results reveal that GPT-3.5-turbo can only detect a small percentage of the bias behaviors in its previously generated code. For instance, GPT-3.5-turbo detects only 18.84\% of the biased codes in the age attribute, while the remaining 81.16\% go undetected. Consequently, when directly requiring GPT-3.5-turbo to remove the bias behaviors in its previously generated code, the CBS only decreases from 23.35\% to 20.36\%. In Scenario 2, however, we also feed the bias results into GPT-3.5-turbo, which further decreases the CBS from 23.35\% to 5.39\%. This is because the biased code that GPT-3.5-turbo fails to detect and mitigate in Scenario 1 is addressed in Scenario 2, as we explicitly inform GPT-3.5-turbo about the specific biases present in the code, which can be detected through the provided test cases.

\mybox{Answer to RQ3: 
Direct prompt engineering strategies have limited effectiveness on bias mitigation in code generation. 
However, with our test analysis feedback, the code biases in all the LLMs under test are significantly reduced. 
For example, the overall CBS decreases from 59.88\% to 4.79\% for GPT-4 with a Chain-of-Thought prompt. The key reason is that LLMs have difficulty detecting bias behaviors in their generated code. However, when we provide feedback to the LLMs, they can then remove the bias from the code that they previously ignored.}

\begin{table}
    \caption{Bias detection results of utilizing LLM to detect bias behaviors for their previously generated code. For each sensitive attribute, we report the accuracy of the GPT-3.5-turbo correctly predicted ratio for the code with the corresponding bias attribute.}
    \centering
    \begin{tabular}{l|*{6}{c}}
    \toprule
         Model & Age & Region & Gender & Education & Occupation & Race \\
         \midrule
        GPT-3.5-turbo & 18.84 & 29.27 &39.47 &12.50  &0.00  &50.00 \\
        \midrule
    \end{tabular}
    \label{tab:LLM_bias_detection_accuracy}
\end{table}

\section{Extended Analysis and Discussion}

\subsection{Is there a trade-off between fairness and performance?}\label{sec:eval:tradeoff}
In traditional machine learning fairness, there is a typical trade-off between fairness and performance~\cite{dutta2020there,barlas2021see,chen2023fairness,chen2022maat,cooper2021emergent,liu2022accuracy}. In this section, we investigate whether such trade-offs also exists in LLMs. Specifically, we estimate the code generation performance of LLMs from the following two aspects. First, the performance of completing our bias sensitive tasks, where we evaluate whether the code generated by LLMs can address tasks based on the prompt requirements. For example, for a task that prompts LLMs to assess the level of employability, we analyze whether the code returns the employability of a person. Second, the general code generation performance in terms of pass@1 of the most widely used HumanEval benchmark~\cite{chen2021evaluating}. For code bias, we focus on the ratio of code with any bias (accumulated from all the protected attributes).

The results are illustrated in \cref{tab:trade-off}, where we observe that the success rate of bias sensitive tasks and pass@1 are generally consistent across different LLMs. However, we observe no trade-offs between bias and these two aspects of code generation performances. In particular, the top three LLMs with the best performance are all GPT models, while GPT-4-turbo and GPT-4 also rank high in terms of bias. The key reason may be that different LLMs are trained with different datasets, and some datasets may contain more biased information than others. Meanwhile, the code generation performance may be affected by several other aspects, such as model training strategies, architecture differences, and optimization techniques.

\begin{table}
    \centering
    % \tiny
    \caption{Trade-off results of bias and code generation performance. Column ``Bias'' shows the absolute number of the biased code and CBS. The following two columns show the number and ratio of successful sensitive coding tasks as well as the pass@1 on the HumanEval benchmark. }
   % \resizebox{\columnwidth}{!}{
    \begin{tabular}{l|ccc}
    \toprule
         Model&Bias&Task completion&pass@1\\
         \midrule
         PALM-2-CodeChat-bison&65 (19.46)&111 (33.23)&43.9\\
         Claude-instant-1&205 (61.38)&183 (54.79)&51.7\\
         GPT-3.5-turbo&145 (43.41)& 211 (63.17)&57.3\\
         GPT-4-turbo&256 (76.65)&210 (62.87)&57.9\\
         GPT-4&203 (60.78)&203 (60.78)&67.0\\
         \bottomrule
    \end{tabular}
    % }
    \label{tab:trade-off}
\end{table}

\subsection{Does the functionality of bias-mitigated code change?}

As shown in \cref{tab:mitigation_with_feedback}, we can observe that the CBS of LLM-generated code after the bias-mitigated process largely decreased compared with the original version, which raises concerns about whether the functionality of LLM-generated code has been changed. Ideally, the code snippets before and after the repair should have similar functionalities regarding inputs with non-sensitive attributes. To demonstrate whether the functionality has been changed, We did a preliminary study on the similarity of bias-mitigated code and initial code based on GraphCodeBERT-Base \cite{Guo2020GraphCodeBERTPC}, where we calculate the of the initial code \cref{tab:big_table} and Scenario 2 generated code \cref{tab:mitigation_with_feedback}. The evaluation results are shown in the \cref{tab:similarity}. We can observe that the CodeBLEU scores range from 0.57 to 0.91. Moreover, we randomly selected 10 code pairs and conducted a manual check. The results show that 7 out of 10 code pairs have similar functionality, while the other three code pairs' functionality has been changed.

\begin{table}[h]
\centering
\caption{Similarity of LLM originally generated code and scenario 2 removed biased code. The evaluation results are calculated by GraphCodeBERT-Base.}
\begin{tabular}{l|ccccc}
\toprule
Model & Zero-Shot & One-Shot & Few-Shot & CoT1 & CoT2 \\
\midrule
palm-2-codechat-bison& 0.86&0.79&0.57&0.87&0.85\\
claude-instant-1& 0.8&0.81&0.73&0.82&0.73\\
gpt-4-turbo-preview& 0.91&0.9&0.76&0.91&0.89\\
gpt-3.5-turbo& 0.76&0.8&0.82&0.8&0.8\\
gpt-4& 0.84&0.85&0.76&0.84&0.84\\
\bottomrule
\end{tabular}
\label{tab:similarity}
\end{table}

\subsection{How do different code generation prompts affect the CBS of LLM-generated code?}

Since minor changes in the prompt may lead to different code generation results, raising concerns about whether the CBS will be subject to change for minor perturbations in prompts. To address this concern, we conducted experiments on five different code generation prompts\footnote{See the prompts in \url{https://github.com/huangd1999/CBS/blob/main/different_prompt_code_generation.py}}.

\paragraph{Semantic similarity for different prompts.}

Before evaluating the CBS of LLM-generated code based on the guidance of different prompts, we first measure the semantic similarity of our five prompts. To measure the semantic similarity of our constructed five prompts, we follow the instructions provided by HuggingFace\footnote{Semantic Similarity: https://huggingface.co/tasks/sentence-similarity\#passage-ranking} to measure the semantic similarity between each pair of prompts. The evaluation results are shown in \cref{tab:prompt_semantic}, where we observe that the semantic similarity between each pair of prompts is larger than 0.8\footnote{If the semantic similarity of two prompts larger than 0.8, we treat them as have the same meaning and goal. See https://docs.llamaindex.ai/en/stable/api\_reference/evaluation/semantic\_similarity/.}, which means that all prompts have the same objective, i.e., they require the LLM to generate code based on the task description.
\begin{table}
\centering
\caption{Semantic similarity for different prompts used in code generation.}
\begin{tabular}{l|rrrrr}
\toprule
Prompt & Prompt1& Prompt2& Prompt3& Prompt4& Prompt5 \\
\midrule
Prompt1 &1.0&0.923&0.907&0.909&0.903 \\
Prompt2 &0.923&1.0&0.915&0.887&0.925 \\
Prompt3 &0.907&0.915&1.0&0.922&0.914 \\
Prompt4 &0.909&0.887&0.922&1.0&0.88 \\
Prompt5 &0.903&0.925&0.914&0.88&1.0 \\
\bottomrule
\end{tabular}
\label{tab:prompt_semantic}
\end{table}

\paragraph{CBS for different prompts generated code.}
Next, we use the provided five different prompts to guide LLMs to generate code based on task description and calculate the CBS of LLM-generated code. The evaluation results are shown in \cref{tab:robustness}, where we can observe differences in GPT-3.5-turbo's output for different prompts. For example, the CBS of age attribute ranges from 21.75\% to 32.93\%, indicating that LLMs may generate code with varying levels of bias depending on the prompt used to guide them. Despite the variations in CBS across different prompts, it is important to note that the CBS remains consistently high for all prompts. For instance, the CBS for the age attribute is consistently above 21\% across all five prompts. This finding suggests that while the specific phrasing of the prompt can influence the extent of bias in the generated code, the overall presence of bias remains a significant concern regardless of the prompt used. 

The impact of prompt phrasing on CBS is further evident when comparing the bias mitigation results of Scenario 1 (Section \ref{sec:without_feedback}) to the direct LLM-generated code. While the results in Scenario 1 show some changes compared to the direct LLM-generated code, they remain similar in some cases, suggesting that the decreased bias in Scenario 1 may be partially attributed to the change in the prompt. However, in Scenario 2, where we provide our bias detection results to the LLM, the CBS is significantly reduced, and in some cases, it decreases to 0. This stark contrast between Scenarios 1 and 2 indicates that our proposed method is effective in mitigating bias, rather than the changes being solely due to the prompt modification. The inclusion of bias detection feedback in Scenario 2 plays a crucial role in guiding the LLM to generate less biased code, demonstrating the effectiveness of our approach in addressing bias in LLM-generated code.

\begin{table}
\centering
\caption{CBS for different prompts generated by GPT-3.5-turbo.}
\begin{tabular}{l|cccccc}
\toprule
Prompt & Age & Region & Gender & Education & Occupation & Race \\
\midrule
Prompt1 & 22.05 & 21.15 & 25.68 & 11.78 & 2.42 & 0.60 \\
Prompt2 & 29.00 & 30.21 & 44.71 & 22.05 & 2.42 & 0.91 \\
Prompt3 & 27.79 & 18.43 & 31.42 & 19.64 & 2.42 & 1.21 \\
Prompt4 & 32.93 & 23.26 & 29.61 & 21.45 & 1.81 & 0.91 \\
Prompt5 & 21.75 & 21.15 & 27.49 & 20.24 & 3.32 & 1.81 \\
\bottomrule
\end{tabular}
\label{tab:robustness}
\end{table}

\subsection{Enhancing Value Pool for Bias Detection}\label{sec:coverage}
demonstrates that our automated bias testing has high precision and recall. However, there are still a few false negatives (FN) due to the uncovered cases in the value pool for the protected attributes. This section explores strategies to enhance the value pool to reduce false negatives. In particular, the limitation observed with age parameters~(i.e., where biases involving ages above 65 are not detected) suggests a gap in our testing scope. To mitigate this, a straightforward solution is to enrich our value pools with a broader range of values, aiming to improve the comprehensiveness of bias detection. Specifically, we add parameter values in ACSIncome, ACSEmployment, and ACSPublicCoverage\footnote{ACSIncome, ACSEmployment, and ACSPublicCoverage provide a range of values for parameters in \cref{tab:attribute}.} \cite{ding2021retiring}, thus improving the coverage of parameter values and addressing the gaps identified in our initial testing framework. The evaluation results are shown in \cref{tab:enriched_tests}, where we can observe that once we add more diverse values to the value pools, the false negative rate decreases to 0. Finally, the recall of automated test case analysis increases from 92\% to 100\%. However, our evaluation results also illustrate that this expansion of the value pool introduces extra overhead for the testing process. Specifically, the testing time increases from 57.15s to 3958.84s. The key reason is that once we increase the value pool for function parameters, the total test cases constructed by \cref{fig:pipeline} \textbf{5a} then largely increase. Considering the large overhead and the small ratio of false negatives, our default strategy does not adopt the large value pool, but users and developers can choose to adapt the value pool to achieve 100\% test recall and precision when necessary.

\begin{table}[h!]
    \centering
    \caption{Evaluation results for TP, FN, FP, and TN when we enrich value pools based on the ACSIncome, ACSEmployment, and ACSPublicCoverage dataset \cite{ding2021retiring}. We also report the testing time in the overhead column.}
    \begin{tabular}{c|ccccc}
    \toprule
& TP&FN&FP&TN&Overhead \\
\midrule
         Original&141 & 12&0&2185 & 57.16s \\
         Enriched&153 & 0&0& 2185 & 3958.84s \\
         \bottomrule
    \end{tabular}
    \label{tab:enriched_tests}
\end{table}

\subsection{Why not use LLM to generate test cases?}

We do not use LLM to generate test cases since LLM-generated test cases are often incorrect, which then requires significant manual efforts to select correct test cases from the generated tests. Besides, the amount of code evaluated in our experiments is extensive (e.g., 334 tasks * 5 models * 5 random generations * 11 scenarios (5 different prompts * 2 scenarios + 1 original code)), and the number of sensitive attributes may range from 1 to 5 for each provided code, which then requires large tokens to generate massive test cases to test the bias in the code. Therefore, we directly use our bias testing framework to construct test cases instead of relying on LLMs for test case generation, which is accurate and efficient.
% }

\subsection{How does temperature affect the CBS of LLM-generated code?}

As shown in \cref{tab:big_table}, the LLM-generated code is different for different execution times, which causes the CBS\_U@5, CBS\_I@5, and CBS to vary across five executions. The key reason for these unstable results is that the temperature setting affects the next token selection. We set the temperature to 1.0 for all experiments. To investigate how temperature affects the bias of the LLM-generated code, we evaluate the CBS of the LLM-generated code at six different temperatures: 0.0, 0.2, 0.4, 0.6, 0.8, and 1.0. The evaluation results are shown in \cref{tab:temperature}, where we can observe that first, the CBS of LLM-generated code varies at different temperatures. The key reason is that the temperature setting in LLMs influences the probability distribution over the next token during generation. Lower temperatures make the model more deterministic, favoring the most likely tokens, while higher temperatures introduce more randomness and encourage the model to explore less likely token choices. Second, we can also observe that even if the temperature is set to 0.0, where the LLM always selects the token with the highest probability in the last layer prediction as the next token during the inference time, the CBS of the LLM-generated code is still not 0, which indicates that during the inference time, LLMs are sometimes trying to generate task with bias behaviors.

\begin{table}
\centering
\caption{CBS for different temperatures generated by GPT-3.5-turbo.}
\begin{tabular}{l|cccccc}
\toprule
Temperature & Age & Region & Gender & Education & Occupation & Race \\
\midrule
0.0&100 (29.07) &44 (12.79) &43 (12.5) &55 (15.99) &17 (4.94) &12 (3.49)\\
0.2&102 (29.65) &40 (11.63) &42 (12.21) &55 (15.99) &22 (6.4) &16 (4.65)\\
0.4&111 (32.27) &42 (12.21) &46 (13.37) &62 (18.02) &20 (5.81) &15 (4.36)\\
0.6&106 (30.81) &40 (11.63) &39 (11.34) &61 (17.73) &16 (4.65) &14 (4.07)\\
0.8&92 (26.74) &37 (10.76) &33 (9.59) &53 (15.41) &25 (7.27) &13 (3.78)\\
1.0&78 (23.35) &46 (13.77) &76 (22.75) &81 (24.25) &5 (1.50) &6 (1.80) \\
\bottomrule
\end{tabular}
\label{tab:temperature}
\end{table}

\subsection{How about the token usage of bias mitigation process?}

As the text window of LLMs is limited, which causes that we can not feed long text into LLMs to mitigate the bias of LLM itself generated code. Then, it is essential to ensure that our bias mitigation approach can effectively handle the programs within the text window. In this section, we measure the token usage of the bias mitigation process and discuss its implications for the scalability and applicability of our approach. We provided the average token usage (input + output) for each LLM used to mitigate bias in the initially generated code for both scenario 1 and scenario 2. As shown in \cref{tab:tokens}, all LLMs for all prompts in both scenarios use less than 4096 tokens (the default token limitation of GPT-3.5, while other LLMs have larger text windows) to complete the tasks. For example,e GPT-3.5-turbo only requires on average 758.15 tokens to finish each task, which are less than the 4096 tokens, which indicates that the size of the programs currently does not affect our approach. Furthermore, as current LLMs are extending their text windows, e.g., GPT-4 has 8K tokens, the newest GPT-3.5 has 16K tokens, and models such as Claude have a 128-200K token window, we believe that the input size in the future would also does not limit the effectiveness of our bias mitigation approach. As LLMs continue to increase their input size capabilities, our method will be able to handle even larger programs without encountering limitations related to token usage.

\begin{table}
\centering
\caption{Average token usage (input + output) for each LLM to mitigate bias code in LLM initial generated code in scenario1 and scenario1. The value of outside/inside the brackets is the scenario1 / scenario2 token usage.}
\begin{tabular}{l|rrrrr}
\toprule
Model & Zero-Shot & One-Shot & Few-Shot & CoT1 & CoT2 \\
\midrule
palm-2-codechat-bison& 742.58 (864.06)&687.71 (927.19)&723.09 (1012.0)&735.3 (977.29)&739.82 (947.18) \\
claude-instant-1& 567.94 (728.0)&557.75 (723.97)&601.61 (786.78)&591.84 (778.65)&553.58 (768.46) \\
gpt-3.5-turbo& 509.91 (817.95)&627.96 (879.73)&758.15 (975.67)&731.45 (938.31)&743.26 (901.21) \\
gpt-4-1106-preview& 900.99 (805.5)&969.46 (888.73)&1036.59 (965.71)&1044.17 (946.55)&1064.04 (912.92) \\
gpt-4& 585.78 (625.74)&633.47 (649.78)&668.82 (710.28)&685.17 (700.04)&703.39 (691.61) \\
\bottomrule
\end{tabular}
\label{tab:tokens}
\end{table}

\section{Threats to Validity}
\subsection{Internal Validity}

The process of creating the code generation prompt dataset involves human judgment, which introduces the possibility of subjective bias in prompt design and may influence the presence or absence of certain biases in the dataset. To mitigate this threat, we ensure consistent and objective prompt creation by employing well-defined operational definitions for each bias type. Additionally, the code generation models may exhibit variations in generating code functions due to inherent randomness and model complexity, potentially impacting the results and introducing internal validity threats. To address this, we carefully control for such variations by running five times for each experiment~(e.g., code generation and bias mitigation) to obtain the average results. Besides, we also utilize CBS\_U@K and CBS\_I@K to decrease the effect of variation for our experiments. These techniques help us reduce the impact of randomness and improve the robustness of our findings. By taking these precautions, we aim to strengthen the internal validity of our research, ensuring the reliability and accuracy of the results obtained from the prompt dataset creation and code generation process.

\subsection{External Validity}
The external validity of our study is subject to the representativeness of the code generation prompt dataset and the generalizability of language models to various code generation tasks. If the dataset does not cover a representative range of potential biases in the code, our findings may lack generalizability to real-world scenarios. To address this concern, we take measures to ensure diversity in the selection of protected attributes and tasks and use the three most widely studied tasks in the fairness literature.

\subsection{Construct Validity}
For code bias evaluation, we rely on automated test case analysis to classify the predominantly generated code functions, providing a more standardized and automated approach. Then, for the code that requires human evaluation due to runtime errors, we have multiple experts to analyze bias types for each code to reduce subjectivity. The construct validity of our study also depends on the effectiveness of the test case analysis result assistant mitigation for the code. If the mitigation approach fails to result in substantial reductions in bias, the validity of our conclusions could be compromised. To mitigate this threat, we conduct comprehensive evaluations to assess five code generation models, test them five times, and report the average results. By doing so, we validate the effectiveness of our mitigation approach and strengthen the construct validity of our research findings.

\section{Related Work}
In this section, we discuss the related work of code generation models and current testing techniques for code generation models.
\subsection{Code Generation Model}

Recently, large language models have been widely used in code generation tasks. Various architectures have been explored in these models, some notable examples being CodeBERT~\cite{codebert}, PLBART \cite{Ahmad2021UnifiedPF}, and CodeGPT~\cite{CERT}. These models are pre-trained on code corpora to develop a deep understanding of code syntax, semantics, and idiomatic constructs. To enhance their comprehension of the complexities in code, some innovative approaches integrate structured representations. For example, CodeT5~\cite{wang2023codet5+} combines the encoder-decoder paradigm with the structural essence of code. These enhancements aim to provide the models with a more fine-grained understanding of code relationships and dependencies beyond just syntactic patterns. A current trend is the construction of large-scale models with billions of parameters, which have illustrated SOTA performance in code generation tasks. Another way is using foundation models~(e.g., PaLM, Claude, ChatGPT, GPT-4) to generate code functions, which have been evaluated for their effectiveness in generating functional code. 

Code generation models have numerous advantages but can also be susceptible to bias that could impact the software they produce. In our study, we carefully investigate this matter, aiming to identify and address biases in automated code generation. Our goal is to enhance the reliability and trustworthiness of the code generated by these models. This highlights the significance of employing bias-aware approaches when utilizing machine assistance in programming tasks. By being mindful of biases, we can ensure more equitable and fair outcomes in the software development process.

\subsection{Testing for Code Generation Model}
To test code generation effectiveness, metrics like BLEU~\cite{bleu} and ROUGE~\cite{rouge} to assess the code's similarity to the canonical solution. Besides, metrics like CodeBLEU~\cite{Ren2020CodeBLEUAM}, METEOR, and CIDEr~\cite{Evtikhiev2022OutOT} refine this analysis, providing a deeper dive into the code's structural and semantic quality. However, while these automated metrics offer quantifiable insights, they often overlook the functional integrity of the code. To address this problem, pass@k has been proposed to bridge this gap. Here, human evaluators assess the code execution accuracy in various test scenarios. Recently, Huang et al. \cite{huang2024effibench} further proposed NET, MU, and TMU to quantify the efficiency of LLM-generated code by measuring execution time and memory usage during the code execution process. Recently, ReCode \cite{Wang2022ReCodeRE} proposes robustness evaluation for code generation models, which focuses on models' resilience, especially under non-ideal or adversarial conditions, which involves introducing perturbations at different granularities and monitoring the model's ability to counteract such disruptions. Different from the above metrics, our research aims to measure the biased behaviors of code generated by LLMs. In addition to our work, there is a simultaneous work conducted during the same period as our research~\cite{liu2023uncovering}, which also focuses on code bias. Although both studies address code bias, Liu et al. \cite{liu2023uncovering}'s focus is limited to code completion. In contrast, our framework concentrates on the broader domain of text-to-code generation, and we also offer practical solutions to reduce biases in AI-generated code. Recently, Ling et al. \cite{ling2024evaluating} primarily investigated social bias inherent in LLMs fine-tuned for code generation tasks. However, Ling et al. \cite{ling2024evaluating} particularly focus in terms of prompt design and testing framework. Our work focuses on both social bias testing and bias mitigation for LLM-generated code.

\section{Conclusion and Future Works}\label{sec:data_availability}
In this work, we propose a code bias testing framework and a large scale empirical study to uncover biases (e.g., age, gender) in code generation models. Based on the framework, we assess current SOTA code generation models, and we observe that all of the tested code generation models sometimes generate biased code functions. We observed that larger language models do not mean fewer code bias behaviors. To mitigate bias in the code generation models, we propose five bias mitigation templates. We release our dataset and source code in~\url{https://github.com/huangd1999/CBS}. In our future work, we will evaluate more code generation models~(e.g., Gemini, Copilot, and CodeX), more bias attributes~(e.g., culture), and more scenarios~(e.g., academy admission).

\section{Acknowledgements}
The work is supported in part by the National Key R\&D Program of China (2022ZD0160201), HK RGC RIF (R7030-22), a Huawei Flagship Research Grant in 2023, HK RGC GRF (Ref: 17208223 \& 17204424), and the HKU-CAS Joint Laboratory for Intelligent System Software.

% \balance
\bibliographystyle{IEEEtran}
\bibliography{acmart-primary/acmart}

\begin{thebibliography}{10}\itemsep=-1pt

\bibitem{andriushchenko2020square}
Maksym Andriushchenko, Francesco Croce, Nicolas Flammarion, and Matthias Hein.
\newblock Square attack: a query-efficient black-box adversarial attack via
  random search.
\newblock In {\em European Conference on Computer Vision}, pages 484--501.
  Springer, 2020.

\bibitem{bernhard2021impact}
R{\'e}mi Bernhard, Pierre-Alain Mo{\"e}llic, Martial Mermillod, Yannick
  Bourrier, Romain Cohendet, Miguel Solinas, and Marina Reyboz.
\newblock Impact of spatial frequency based constraints on adversarial
  robustness.
\newblock In {\em 2021 International Joint Conference on Neural Networks
  (IJCNN)}, pages 1--8. IEEE, 2021.

\bibitem{brendel2017decision}
Wieland Brendel, Jonas Rauber, and Matthias Bethge.
\newblock Decision-based adversarial attacks: Reliable attacks against
  black-box machine learning models.
\newblock {\em arXiv preprint arXiv:1712.04248}, 2017.

\bibitem{carlini2017towards}
Nicholas Carlini and David Wagner.
\newblock Towards evaluating the robustness of neural networks.
\newblock In {\em 2017 ieee symposium on security and privacy (sp)}, pages
  39--57. Ieee, 2017.

\bibitem{croce2020minimally}
Francesco Croce and Matthias Hein.
\newblock Minimally distorted adversarial examples with a fast adaptive
  boundary attack.
\newblock In {\em International Conference on Machine Learning}, pages
  2196--2205. PMLR, 2020.

\bibitem{croce2020reliable}
Francesco Croce and Matthias Hein.
\newblock Reliable evaluation of adversarial robustness with an ensemble of
  diverse parameter-free attacks.
\newblock In {\em International conference on machine learning}, pages
  2206--2216. PMLR, 2020.

\bibitem{cui2021learnable}
Jiequan Cui, Shu Liu, Liwei Wang, and Jiaya Jia.
\newblock Learnable boundary guided adversarial training.
\newblock In {\em Proceedings of the IEEE/CVF international conference on
  computer vision}, pages 15721--15730, 2021.

\bibitem{deng2020analysis}
Yao Deng, Xi Zheng, Tianyi Zhang, Chen Chen, Guannan Lou, and Miryung Kim.
\newblock An analysis of adversarial attacks and defenses on autonomous driving
  models.
\newblock In {\em 2020 IEEE international conference on pervasive computing and
  communications (PerCom)}, pages 1--10. IEEE, 2020.

\bibitem{dong2019evading}
Yinpeng Dong, Tianyu Pang, Hang Su, and Jun Zhu.
\newblock Evading defenses to transferable adversarial examples by
  translation-invariant attacks.
\newblock In {\em Proceedings of the IEEE/CVF Conference on Computer Vision and
  Pattern Recognition}, pages 4312--4321, 2019.

\bibitem{finlayson2019adversarial}
Samuel~G Finlayson, John~D Bowers, Joichi Ito, Jonathan~L Zittrain, Andrew~L
  Beam, and Isaac~S Kohane.
\newblock Adversarial attacks on medical machine learning.
\newblock {\em Science}, 363(6433):1287--1289, 2019.

\bibitem{geirhos2018imagenet}
Robert Geirhos, Patricia Rubisch, Claudio Michaelis, Matthias Bethge, Felix~A
  Wichmann, and Wieland Brendel.
\newblock Imagenet-trained cnns are biased towards texture; increasing shape
  bias improves accuracy and robustness.
\newblock {\em arXiv preprint arXiv:1811.12231}, 2018.

\bibitem{goodfellow2014explaining}
Ian~J Goodfellow, Jonathon Shlens, and Christian Szegedy.
\newblock Explaining and harnessing adversarial examples.
\newblock {\em arXiv preprint arXiv:1412.6572}, 2014.

\bibitem{gowal2020uncovering}
Sven Gowal, Chongli Qin, Jonathan Uesato, Timothy Mann, and Pushmeet Kohli.
\newblock Uncovering the limits of adversarial training against norm-bounded
  adversarial examples.
\newblock {\em arXiv preprint arXiv:2010.03593}, 2020.

\bibitem{gurel2021knowledge}
Nezihe~Merve G{\"u}rel, Xiangyu Qi, Luka Rimanic, Ce Zhang, and Bo Li.
\newblock Knowledge enhanced machine learning pipeline against diverse
  adversarial attacks.
\newblock In {\em International Conference on Machine Learning}, pages
  3976--3987. PMLR, 2021.

\bibitem{he2016deep}
Kaiming He, Xiangyu Zhang, Shaoqing Ren, and Jian Sun.
\newblock Deep residual learning for image recognition.
\newblock In {\em Proceedings of the IEEE conference on computer vision and
  pattern recognition}, pages 770--778, 2016.

\bibitem{huang2022frequency}
Binxiao Huang, Chaofan Tao, Rui Lin, and Ngai Wong.
\newblock Frequency regularization for improving adversarial robustness.
\newblock {\em arXiv preprint arXiv:2212.12732}, 2022.

\bibitem{ilyas2018black}
Andrew Ilyas, Logan Engstrom, Anish Athalye, and Jessy Lin.
\newblock Black-box adversarial attacks with limited queries and information.
\newblock In {\em International Conference on Machine Learning}, pages
  2137--2146. PMLR, 2018.

\bibitem{izmailov2018averaging}
Pavel Izmailov, Dmitrii Podoprikhin, Timur Garipov, Dmitry Vetrov, and
  Andrew~Gordon Wilson.
\newblock Averaging weights leads to wider optima and better generalization.
\newblock {\em arXiv preprint arXiv:1803.05407}, 2018.

\bibitem{jia2022adversarial}
Xiaojun Jia, Yong Zhang, Baoyuan Wu, Ke Ma, Jue Wang, and Xiaochun Cao.
\newblock Las-at: adversarial training with learnable attack strategy.
\newblock In {\em Proceedings of the IEEE/CVF Conference on Computer Vision and
  Pattern Recognition}, pages 13398--13408, 2022.

\bibitem{krizhevsky2009learning}
Alex Krizhevsky, Geoffrey Hinton, et~al.
\newblock Learning multiple layers of features from tiny images.
\newblock 2009.

\bibitem{kurakin2016adversarial}
Alexey Kurakin, Ian Goodfellow, and Samy Bengio.
\newblock Adversarial machine learning at scale.
\newblock {\em arXiv preprint arXiv:1611.01236}, 2016.

\bibitem{madry2017towards}
Aleksander Madry, Aleksandar Makelov, Ludwig Schmidt, Dimitris Tsipras, and
  Adrian Vladu.
\newblock Towards deep learning models resistant to adversarial attacks.
\newblock {\em arXiv preprint arXiv:1706.06083}, 2017.

\bibitem{maiya2021frequency}
Shishira~R Maiya, Max Ehrlich, Vatsal Agarwal, Ser-Nam Lim, Tom Goldstein, and
  Abhinav Shrivastava.
\newblock A frequency perspective of adversarial robustness.
\newblock {\em arXiv preprint arXiv:2111.00861}, 2021.

\bibitem{moosavi2016deepfool}
Seyed-Mohsen Moosavi-Dezfooli, Alhussein Fawzi, and Pascal Frossard.
\newblock Deepfool: a simple and accurate method to fool deep neural networks.
\newblock In {\em Proceedings of the IEEE conference on computer vision and
  pattern recognition}, pages 2574--2582, 2016.

\bibitem{park2022vision}
Namuk Park and Songkuk Kim.
\newblock How do vision transformers work?
\newblock {\em arXiv preprint arXiv:2202.06709}, 2022.

\bibitem{qin2019adversarial}
Chongli Qin, James Martens, Sven Gowal, Dilip Krishnan, Krishnamurthy
  Dvijotham, Alhussein Fawzi, Soham De, Robert Stanforth, and Pushmeet Kohli.
\newblock Adversarial robustness through local linearization.
\newblock {\em Advances in Neural Information Processing Systems}, 32, 2019.

\bibitem{salzmann2021learning}
Mathieu Salzmann et~al.
\newblock Learning transferable adversarial perturbations.
\newblock {\em Advances in Neural Information Processing Systems},
  34:13950--13962, 2021.

\bibitem{shafahi2019adversarial}
Ali Shafahi, Mahyar Najibi, Mohammad~Amin Ghiasi, Zheng Xu, John Dickerson,
  Christoph Studer, Larry~S Davis, Gavin Taylor, and Tom Goldstein.
\newblock Adversarial training for free!
\newblock {\em Advances in Neural Information Processing Systems}, 32, 2019.

\bibitem{shao2021adversarial}
Rulin Shao, Zhouxing Shi, Jinfeng Yi, Pin-Yu Chen, and Cho-Jui Hsieh.
\newblock On the adversarial robustness of vision transformers.
\newblock {\em arXiv preprint arXiv:2103.15670}, 2021.

\bibitem{szegedy2013intriguing}
Christian Szegedy, Wojciech Zaremba, Ilya Sutskever, Joan Bruna, Dumitru Erhan,
  Ian Goodfellow, and Rob Fergus.
\newblock Intriguing properties of neural networks.
\newblock {\em arXiv preprint arXiv:1312.6199}, 2013.

\bibitem{wang2022anti}
Peihao Wang, Wenqing Zheng, Tianlong Chen, and Zhangyang Wang.
\newblock Anti-oversmoothing in deep vision transformers via the fourier domain
  analysis: From theory to practice.
\newblock {\em arXiv preprint arXiv:2203.05962}, 2022.

\bibitem{wang2020improving}
Yisen Wang, Difan Zou, Jinfeng Yi, James Bailey, Xingjun Ma, and Quanquan Gu.
\newblock Improving adversarial robustness requires revisiting misclassified
  examples.
\newblock In {\em International Conference on Learning Representations}, 2020.

\bibitem{wong2020fast}
Eric Wong, Leslie Rice, and J~Zico Kolter.
\newblock Fast is better than free: Revisiting adversarial training.
\newblock {\em arXiv preprint arXiv:2001.03994}, 2020.

\bibitem{wu2020adversarial}
Dongxian Wu, Shu-Tao Xia, and Yisen Wang.
\newblock Adversarial weight perturbation helps robust generalization.
\newblock {\em Advances in Neural Information Processing Systems},
  33:2958--2969, 2020.

\bibitem{xie2019improving}
Cihang Xie, Zhishuai Zhang, Yuyin Zhou, Song Bai, Jianyu Wang, Zhou Ren, and
  Alan~L Yuille.
\newblock Improving transferability of adversarial examples with input
  diversity.
\newblock In {\em Proceedings of the IEEE/CVF Conference on Computer Vision and
  Pattern Recognition}, pages 2730--2739, 2019.

\bibitem{xu2022overview}
Zhi-Qin~John Xu, Yaoyu Zhang, and Tao Luo.
\newblock Overview frequency principle/spectral bias in deep learning.
\newblock {\em arXiv preprint arXiv:2201.07395}, 2022.

\bibitem{yue2021robust}
Jiutao Yue, Haofeng Li, Pengxu Wei, Guanbin Li, and Liang Lin.
\newblock Robust real-world image super-resolution against adversarial attacks.
\newblock In {\em Proceedings of the 29th ACM International Conference on
  Multimedia}, pages 5148--5157, 2021.

\bibitem{zagoruyko2016wide}
Sergey Zagoruyko and Nikos Komodakis.
\newblock Wide residual networks.
\newblock {\em arXiv preprint arXiv:1605.07146}, 2016.

\bibitem{zhang2019theoretically}
Hongyang Zhang, Yaodong Yu, Jiantao Jiao, Eric Xing, Laurent El~Ghaoui, and
  Michael Jordan.
\newblock Theoretically principled trade-off between robustness and accuracy.
\newblock In {\em International conference on machine learning}, pages
  7472--7482. PMLR, 2019.

\bibitem{zhang2020attacks}
Jingfeng Zhang, Xilie Xu, Bo Han, Gang Niu, Lizhen Cui, Masashi Sugiyama, and
  Mohan Kankanhalli.
\newblock Attacks which do not kill training make adversarial learning
  stronger.
\newblock In {\em International conference on machine learning}, pages
  11278--11287. PMLR, 2020.

\bibitem{zhang2020geometry}
Jingfeng Zhang, Jianing Zhu, Gang Niu, Bo Han, Masashi Sugiyama, and Mohan
  Kankanhalli.
\newblock Geometry-aware instance-reweighted adversarial training.
\newblock {\em arXiv preprint arXiv:2010.01736}, 2020.

\bibitem{zhang2019interpreting}
Tianyuan Zhang and Zhanxing Zhu.
\newblock Interpreting adversarially trained convolutional neural networks.
\newblock In {\em International conference on machine learning}, pages
  7502--7511. PMLR, 2019.

\bibitem{zhang2019adversarial}
Zhendong Zhang, Cheolkon Jung, and Xiaolong Liang.
\newblock Adversarial defense by suppressing high-frequency components.
\newblock {\em arXiv preprint arXiv:1908.06566}, 2019.

\end{thebibliography}


\begin{thebibliography}{10}
\providecommand{\url}[1]{#1}
\csname url@samestyle\endcsname
\providecommand{\newblock}{\relax}
\providecommand{\bibinfo}[2]{#2}
\providecommand{\BIBentrySTDinterwordspacing}{\spaceskip=0pt\relax}
\providecommand{\BIBentryALTinterwordstretchfactor}{4}
\providecommand{\BIBentryALTinterwordspacing}{\spaceskip=\fontdimen2\font plus
\BIBentryALTinterwordstretchfactor\fontdimen3\font minus \fontdimen4\font\relax}
\providecommand{\BIBforeignlanguage}[2]{{%
\expandafter\ifx\csname l@#1\endcsname\relax
\typeout{** WARNING: IEEEtran.bst: No hyphenation pattern has been}%
\typeout{** loaded for the language `#1'. Using the pattern for}%
\typeout{** the default language instead.}%
\else
\language=\csname l@#1\endcsname
\fi
#2}}
\providecommand{\BIBdecl}{\relax}
\BIBdecl

\bibitem{chen2021evaluating}
M.~Chen, J.~Tworek, H.~Jun, Q.~Yuan, H.~P. d.~O. Pinto, J.~Kaplan, H.~Edwards, Y.~Burda, N.~Joseph, G.~Brockman \emph{et~al.}, ``Evaluating large language models trained on code,'' \emph{arXiv preprint arXiv:2107.03374}, 2021.

\bibitem{OpenAI2023GPT4TR}
OpenAI, ``Gpt-4 technical report,'' \emph{ArXiv}, vol. abs/2303.08774, 2023.

\bibitem{Liu2023UncoveringAQ}
Y.~Liu, X.~Chen, Y.~Gao, Z.~Su, F.~Zhang, D.~Zan, J.-G. Lou, P.-Y. Chen, and T.-Y. Ho, ``Uncovering and quantifying social biases in code generation,'' vol.~36, pp. 2368--2380, 2023.

\bibitem{mukherjee2014role}
A.~N. Mukherjee, S.~Bhattacharyya, and R.~Bera, ``Role of information technology in human resource management of sme: A study on the use of applicant tracking system,'' \emph{IBMRD's Journal of Management \& Research}, pp. 1--22, 2014.

\bibitem{ahmad2009smart}
N.~Ahmad and A.~N. Abd~Alla, ``Smart evaluation for job vacancy application system,'' in \emph{2009 Second International Conference on the Applications of Digital Information and Web Technologies}.\hskip 1em plus 0.5em minus 0.4em\relax IEEE, 2009, pp. 452--455.

\bibitem{thakur2023language}
H.~Thakur, A.~Jain, P.~Vaddamanu, P.~P. Liang, and L.-P. Morency, ``Language models get a gender makeover: Mitigating gender bias with few-shot data interventions,'' \emph{arXiv preprint arXiv:2306.04597}, 2023.

\bibitem{ungless2022robust}
E.~L. Ungless, A.~Rafferty, H.~Nag, and B.~Ross, ``A robust bias mitigation procedure based on the stereotype content model,'' \emph{arXiv preprint arXiv:2210.14552}, 2022.

\bibitem{lee2023kosbi}
H.~Lee, S.~Hong, J.~Park, T.~Kim, G.~Kim, and J.-W. Ha, ``Kosbi: A dataset for mitigating social bias risks towards safer large language model application,'' \emph{arXiv preprint arXiv:2305.17701}, 2023.

\bibitem{barikeri2021redditbias}
S.~Barikeri, A.~Lauscher, I.~Vuli{\'c}, and G.~Glava{\v{s}}, ``Redditbias: A real-world resource for bias evaluation and debiasing of conversational language models,'' \emph{arXiv preprint arXiv:2106.03521}, 2021.

\bibitem{felkner2023winoqueer}
V.~K. Felkner, H.-C.~H. Chang, E.~Jang, and J.~May, ``Winoqueer: A community-in-the-loop benchmark for anti-lgbtq+ bias in large language models,'' \emph{arXiv preprint arXiv:2306.15087}, 2023.

\bibitem{fleisig2022mitigating}
E.~Fleisig and C.~Fellbaum, ``Mitigating gender bias in machine translation through adversarial learning,'' \emph{arXiv preprint arXiv:2203.10675}, 2022.

\bibitem{biswas23fairify}
S.~Biswas and H.~Rajan, ``Fairify: Fairness verification of neural networks,'' in \emph{ICSE'2023: The 45th International Conference on Software Engineering}, May 14-May 20 2023.

\bibitem{gohar23fairness}
S.~B. Usman~Gohar and H.~Rajan, ``Towards understanding fairness and its composition in ensemble machine learning,'' in \emph{ICSE'2023: The 45th International Conference on Software Engineering}, May 14-May 20 2023.

\bibitem{biswas20machine}
S.~Biswas and H.~Rajan, ``Do the machine learning models on a crowd sourced platform exhibit bias? an empirical study on model fairness,'' in \emph{ESEC/FSE'2020: The 28th ACM Joint European Software Engineering Conference and Symposium on the Foundations of Software Engineering}, November 8-November 13, 2020 2020.

\bibitem{2020t5}
\BIBentryALTinterwordspacing
C.~Raffel, N.~Shazeer, A.~Roberts, K.~Lee, S.~Narang, M.~Matena, Y.~Zhou, W.~Li, and P.~J. Liu, ``Exploring the limits of transfer learning with a unified text-to-text transformer,'' \emph{Journal of Machine Learning Research}, vol.~21, no. 140, pp. 1--67, 2020. [Online]. Available: \url{http://jmlr.org/papers/v21/20-074.html}
\BIBentrySTDinterwordspacing

\bibitem{wang2023codet5+}
Y.~Wang, H.~Le, A.~D. Gotmare, N.~D. Bui, J.~Li, and S.~C. Hoi, ``Codet5+: Open code large language models for code understanding and generation,'' \emph{arXiv preprint arXiv:2305.07922}, 2023.

\bibitem{ouyang2023llm}
S.~Ouyang, J.~M. Zhang, M.~Harman, and M.~Wang, ``Llm is like a box of chocolates: the non-determinism of chatgpt in code generation,'' \emph{arXiv preprint arXiv:2308.02828}, 2023.

\bibitem{Wang2022ReCodeRE}
\BIBentryALTinterwordspacing
S.~Wang, Z.~Li, H.~Qian, C.~Yang, Z.~Wang, M.~Shang, V.~Kumar, S.~Tan, B.~Ray, P.~Bhatia, R.~Nallapati, M.~K. Ramanathan, D.~Roth, and B.~Xiang, ``Recode: Robustness evaluation of code generation models,'' \emph{ArXiv}, vol. abs/2212.10264, 2022. [Online]. Available: \url{https://api.semanticscholar.org/CorpusID:254877229}
\BIBentrySTDinterwordspacing

\bibitem{alayrac2022flamingo}
J.-B. Alayrac, J.~Donahue, P.~Luc, A.~Miech, I.~Barr, Y.~Hasson, K.~Lenc, A.~Mensch, K.~Millican, M.~Reynolds \emph{et~al.}, ``Flamingo: a visual language model for few-shot learning,'' \emph{Advances in Neural Information Processing Systems}, vol.~35, pp. 23\,716--23\,736, 2022.

\bibitem{izacard2022few}
G.~Izacard, P.~Lewis, M.~Lomeli, L.~Hosseini, F.~Petroni, T.~Schick, J.~Dwivedi-Yu, A.~Joulin, S.~Riedel, and E.~Grave, ``Few-shot learning with retrieval augmented language models,'' \emph{arXiv preprint arXiv:2208.03299}, 2022.

\bibitem{tunstall2022efficient}
L.~Tunstall, N.~Reimers, U.~E.~S. Jo, L.~Bates, D.~Korat, M.~Wasserblat, and O.~Pereg, ``Efficient few-shot learning without prompts,'' \emph{arXiv preprint arXiv:2209.11055}, 2022.

\bibitem{wei2022chain}
J.~Wei, X.~Wang, D.~Schuurmans, M.~Bosma, F.~Xia, E.~Chi, Q.~V. Le, D.~Zhou \emph{et~al.}, ``Chain-of-thought prompting elicits reasoning in large language models,'' \emph{Advances in Neural Information Processing Systems}, vol.~35, pp. 24\,824--24\,837, 2022.

\bibitem{madaan2022text}
A.~Madaan and A.~Yazdanbakhsh, ``Text and patterns: For effective chain of thought, it takes two to tango,'' \emph{arXiv preprint arXiv:2209.07686}, 2022.

\bibitem{wang2022self}
X.~Wang, J.~Wei, D.~Schuurmans, Q.~Le, E.~Chi, S.~Narang, A.~Chowdhery, and D.~Zhou, ``Self-consistency improves chain of thought reasoning in language models,'' \emph{arXiv preprint arXiv:2203.11171}, 2022.

\bibitem{chu2023survey}
Z.~Chu, J.~Chen, Q.~Chen, W.~Yu, T.~He, H.~Wang, W.~Peng, M.~Liu, B.~Qin, and T.~Liu, ``A survey of chain of thought reasoning: Advances, frontiers and future,'' \emph{arXiv preprint arXiv:2309.15402}, 2023.

\bibitem{huang2023codecot}
D.~Huang, Q.~Bu, and H.~Cui, ``Codecot and beyond: Learning to program and test like a developer,'' \emph{arXiv preprint arXiv:2308.08784}, 2023.

\bibitem{huang2023agentcoder}
D.~Huang, Q.~Bu, J.~M. Zhang, M.~Luck, and H.~Cui, ``Agentcoder: Multi-agent-based code generation with iterative testing and optimisation,'' \emph{arXiv preprint arXiv:2312.13010}, 2023.

\bibitem{obrien24prompt}
D.~OBrien, S.~Biswas, S.~Imtiaz, R.~Abdalkareem, E.~Shihab, and H.~Rajan, ``Are prompt engineering and todo comments friends or foes? an evaluation on github copilot,'' in \emph{ICSE'2024: The 46th International Conference on Software Engineering}, April 14-April 20 2024.

\bibitem{chen2023comprehensive}
Z.~Chen, J.~M. Zhang, F.~Sarro, and M.~Harman, ``A comprehensive empirical study of bias mitigation methods for machine learning classifiers,'' \emph{ACM Transactions on Software Engineering and Methodology}, vol.~32, no.~4, pp. 1--30, 2023.

\bibitem{ding2021retiring}
F.~Ding, M.~Hardt, J.~Miller, and L.~Schmidt, ``Retiring adult: New datasets for fair machine learning,'' \emph{Advances in neural information processing systems}, vol.~34, pp. 6478--6490, 2021.

\bibitem{le2022survey}
T.~Le~Quy, A.~Roy, V.~Iosifidis, W.~Zhang, and E.~Ntoutsi, ``A survey on datasets for fairness-aware machine learning,'' \emph{Wiley Interdisciplinary Reviews: Data Mining and Knowledge Discovery}, vol.~12, no.~3, p. e1452, 2022.

\bibitem{friedler2019comparative}
S.~A. Friedler, C.~Scheidegger, S.~Venkatasubramanian, S.~Choudhary, E.~P. Hamilton, and D.~Roth, ``A comparative study of fairness-enhancing interventions in machine learning,'' in \emph{Proceedings of the conference on fairness, accountability, and transparency}, 2019, pp. 329--338.

\bibitem{besse2022survey}
P.~Besse, E.~del Barrio, P.~Gordaliza, J.-M. Loubes, and L.~Risser, ``A survey of bias in machine learning through the prism of statistical parity,'' \emph{The American Statistician}, vol.~76, no.~2, pp. 188--198, 2022.

\bibitem{kang2021multifair}
J.~Kang, T.~Xie, X.~Wu, R.~Maciejewski, and H.~Tong, ``Multifair: Multi-group fairness in machine learning,'' \emph{arXiv preprint arXiv:2105.11069}, 2021.

\bibitem{mehrabi2021survey}
N.~Mehrabi, F.~Morstatter, N.~Saxena, K.~Lerman, and A.~Galstyan, ``A survey on bias and fairness in machine learning,'' \emph{ACM computing surveys (CSUR)}, vol.~54, no.~6, pp. 1--35, 2021.

\bibitem{kearns2019empirical}
M.~Kearns, S.~Neel, A.~Roth, and Z.~S. Wu, ``An empirical study of rich subgroup fairness for machine learning,'' in \emph{Proceedings of the conference on fairness, accountability, and transparency}, 2019, pp. 100--109.

\bibitem{komiyama2017two}
J.~Komiyama and H.~Shimao, ``Two-stage algorithm for fairness-aware machine learning,'' \emph{arXiv preprint arXiv:1710.04924}, 2017.

\bibitem{xia2022summer}
F.~Xia, T.~Guo, X.~Bai, A.~Shatte, Z.~Liu, and J.~Tang, ``Summer: Bias-aware prediction of graduate employment based on educational big data,'' \emph{ACM/IMS Transactions on Data Science (TDS)}, vol.~2, no.~4, pp. 1--24, 2022.

\bibitem{papadaki2022federated}
A.~Papadaki, N.~Martinez, M.~A. Bertran, G.~Sapiro, and M.~R. Rodrigues, ``Federated fairness without access to demographics,'' in \emph{Workshop on Federated Learning: Recent Advances and New Challenges (in Conjunction with NeurIPS 2022)}, 2022.

\bibitem{han2023retiring}
X.~Han, Z.~Jiang, H.~Jin, Z.~Liu, N.~Zou, Q.~Wang, and X.~Hu, ``Retiring dp: New distribution-level metrics for demographic parity,'' \emph{Transactions on Machine Learning Research}, 2023.

\bibitem{papadaki2022minimax}
A.~Papadaki, N.~Martinez, M.~Bertran, G.~Sapiro, and M.~Rodrigues, ``Minimax demographic group fairness in federated learning,'' in \emph{Proceedings of the 2022 ACM Conference on Fairness, Accountability, and Transparency}, 2022, pp. 142--159.

\bibitem{mougan2023demographic}
C.~Mougan, L.~State, A.~Ferrara, S.~Ruggieri, and S.~Staab, ``Demographic parity inspector: Fairness audits via the explanation space,'' \emph{arXiv preprint arXiv:2303.08040}, 2023.

\bibitem{de2023empirical}
A.~S. de~Oliveira, C.~Kaplan, K.~Mallat, and T.~Chakraborty, ``An empirical analysis of fairness notions under differential privacy,'' \emph{arXiv preprint arXiv:2302.02910}, 2023.

\bibitem{ferry2023addresing}
J.~Ferry, ``Addresing interpretability fairness \& privacy in machine learning through combinatorial optimization methods,'' Ph.D. dissertation, Universit{\'e} Paul Sabatier-Toulouse III, 2023.

\bibitem{wang2022towards}
A.~Wang, V.~V. Ramaswamy, and O.~Russakovsky, ``Towards intersectionality in machine learning: Including more identities, handling underrepresentation, and performing evaluation,'' in \emph{Proceedings of the 2022 ACM Conference on Fairness, Accountability, and Transparency}, 2022, pp. 336--349.

\bibitem{sattigeri2022fair}
P.~Sattigeri, S.~Ghosh, I.~Padhi, P.~Dognin, and K.~R. Varshney, ``Fair infinitesimal jackknife: Mitigating the influence of biased training data points without refitting,'' \emph{Advances in Neural Information Processing Systems}, vol.~35, pp. 35\,894--35\,906, 2022.

\bibitem{gardner2022subgroup}
J.~Gardner, Z.~Popovic, and L.~Schmidt, ``Subgroup robustness grows on trees: An empirical baseline investigation,'' \emph{Advances in Neural Information Processing Systems}, vol.~35, pp. 9939--9954, 2022.

\bibitem{ferry2023exploiting}
J.~Ferry, U.~A{\"\i}vodji, S.~Gambs, M.-J. Huguet, and M.~Siala, ``Exploiting fairness to enhance sensitive attributes reconstruction,'' in \emph{2023 IEEE Conference on Secure and Trustworthy Machine Learning (SaTML)}.\hskip 1em plus 0.5em minus 0.4em\relax IEEE, 2023, pp. 18--41.

\bibitem{cruz2023unprocessing}
A.~F. Cruz and M.~Hardt, ``Unprocessing seven years of algorithmic fairness,'' \emph{arXiv preprint arXiv:2306.07261}, 2023.

\bibitem{alvarez2023domain}
J.~M. Alvarez, K.~M. Scott, B.~Berendt, and S.~Ruggieri, ``Domain adaptive decision trees: Implications for accuracy and fairness,'' in \emph{Proceedings of the 2023 ACM Conference on Fairness, Accountability, and Transparency}, 2023, pp. 423--433.

\bibitem{cruz2022fairgbm}
A.~F. Cruz, C.~Bel{\'e}m, S.~Jesus, J.~Bravo, P.~Saleiro, and P.~Bizarro, ``Fairgbm: Gradient boosting with fairness constraints,'' \emph{arXiv preprint arXiv:2209.07850}, 2022.

\bibitem{bharti2024estimating}
B.~Bharti, P.~Yi, and J.~Sulam, ``Estimating and controlling for equalized odds via sensitive attribute predictors,'' \emph{Advances in Neural Information Processing Systems}, vol.~36, 2024.

\bibitem{simson2023using}
J.~Simson, F.~Pfisterer, and C.~Kern, ``Using multiverse analysis to evaluate the influence of model design decisions on algorithmic fairness,'' in \emph{HHAI 2023: Augmenting Human Intellect}.\hskip 1em plus 0.5em minus 0.4em\relax IOS Press, 2023, pp. 382--384.

\bibitem{nguyen23fix}
G.~Nguyen, S.~Biswas, and H.~Rajan, ``Fix fairness, don't ruin accuracy: Performance aware fairness repair using automl,'' in \emph{ESEC/FSE'2023: The 31st ACM Joint European Software Engineering Conference and Symposium on the Foundations of Software Engineering}, December 3-9, 2023 2023.

\bibitem{andreeva2004impact}
G.~Andreeva, J.~Ansell, and J.~Crook, ``Impact of anti-discrimination laws on credit scoring,'' \emph{Journal of Financial Services Marketing}, vol.~9, pp. 22--33, 2004.

\bibitem{chouldechova2018frontiers}
A.~Chouldechova and A.~Roth, ``The frontiers of fairness in machine learning,'' \emph{arXiv preprint arXiv:1810.08810}, 2018.

\bibitem{tizpaz2022fairness}
S.~Tizpaz-Niari, A.~Kumar, G.~Tan, and A.~Trivedi, ``Fairness-aware configuration of machine learning libraries,'' in \emph{Proceedings of the 44th International Conference on Software Engineering}, 2022, pp. 909--920.

\bibitem{chang2021privacy}
H.~Chang and R.~Shokri, ``On the privacy risks of algorithmic fairness,'' in \emph{2021 IEEE European Symposium on Security and Privacy (EuroS\&P)}.\hskip 1em plus 0.5em minus 0.4em\relax IEEE, 2021, pp. 292--303.

\bibitem{corbett2018measure}
S.~Corbett-Davies and S.~Goel, ``The measure and mismeasure of fairness: A critical review of fair machine learning,'' \emph{arXiv preprint arXiv:1808.00023}, 2018.

\bibitem{chen2024fairness}
Z.~Chen, J.~M. Zhang, F.~Sarro, and M.~Harman, ``Fairness improvement with multiple protected attributes: How far are we?''\hskip 1em plus 0.5em minus 0.4em\relax IEEE/ACM, 2024.

\bibitem{salewski2023context}
L.~Salewski, S.~Alaniz, I.~Rio-Torto, E.~Schulz, and Z.~Akata, ``In-context impersonation reveals large language models' strengths and biases,'' \emph{arXiv preprint arXiv:2305.14930}, 2023.

\bibitem{wang2023large}
P.~Wang, L.~Li, L.~Chen, D.~Zhu, B.~Lin, Y.~Cao, Q.~Liu, T.~Liu, and Z.~Sui, ``Large language models are not fair evaluators,'' \emph{arXiv preprint arXiv:2305.17926}, 2023.

\bibitem{yu2023large}
Y.~Yu, Y.~Zhuang, J.~Zhang, Y.~Meng, A.~Ratner, R.~Krishna, J.~Shen, and C.~Zhang, ``Large language model as attributed training data generator: A tale of diversity and bias,'' \emph{arXiv preprint arXiv:2306.15895}, 2023.

\bibitem{hernandez2019bargaining}
M.~Hernandez, D.~R. Avery, S.~D. Volpone, and C.~R. Kaiser, ``Bargaining while black: The role of race in salary negotiations.'' \emph{Journal of Applied Psychology}, vol. 104, no.~4, p. 581, 2019.

\bibitem{arceo2022gender}
E.~O. Arceo-Gomez, R.~M. Campos-Vazquez, R.~Y. Badillo, and S.~Lopez-Araiza, ``Gender stereotypes in job advertisements: What do they imply for the gender salary gap?'' \emph{Journal of Labor Research}, vol.~43, no.~1, pp. 65--102, 2022.

\bibitem{taylor2020salary}
L.~L. Taylor, J.~N. Lahey, M.~I. Beck, and J.~E. Froyd, ``How to do a salary equity study: With an illustrative example from higher education,'' \emph{Public personnel management}, vol.~49, no.~1, pp. 57--82, 2020.

\bibitem{platteau2021cognitive}
J.-P. Platteau and D.~U. Ontiveros, ``Cognitive bias in insurance: evidence from a health scheme in india,'' \emph{World Development}, vol. 144, p. 105498, 2021.

\bibitem{adult_income}
``Adult income dataset,'' \url{www.kaggle.com/datasets/wenruliu/adult-income-dataset}, 2023, accessed on August 1, 2023.

\bibitem{employee}
``Employee dataset,'' \url{www.kaggle.com/datasets/tawfikelmetwally/employee-dataset}, 2023, accessed on August 1, 2023.

\bibitem{insurance}
``Us health insurance dataset,'' \url{www.kaggle.com/datasets/teertha/ushealthinsurancedataset}, 2023, accessed on August 1, 2023.

\bibitem{Mehrabi2019ASO}
\BIBentryALTinterwordspacing
N.~Mehrabi, F.~Morstatter, N.~A. Saxena, K.~Lerman, and A.~G. Galstyan, ``A survey on bias and fairness in machine learning,'' \emph{ACM Computing Surveys (CSUR)}, vol.~54, pp. 1 -- 35, 2019. [Online]. Available: \url{https://api.semanticscholar.org/CorpusID:201666566}
\BIBentrySTDinterwordspacing

\bibitem{dutta2020there}
S.~Dutta, D.~Wei, H.~Yueksel, P.-Y. Chen, S.~Liu, and K.~Varshney, ``Is there a trade-off between fairness and accuracy? a perspective using mismatched hypothesis testing,'' in \emph{International conference on machine learning}.\hskip 1em plus 0.5em minus 0.4em\relax PMLR, 2020, pp. 2803--2813.

\bibitem{barlas2021see}
P.~Barlas, K.~Kyriakou, O.~Guest, S.~Kleanthous, and J.~Otterbacher, ``To" see" is to stereotype: Image tagging algorithms, gender recognition, and the accuracy-fairness trade-off,'' \emph{Proceedings of the ACM on Human-Computer Interaction}, vol.~4, no. CSCW3, pp. 1--31, 2021.

\bibitem{chen2023fairness}
Z.~Chen, J.~Zhang, F.~Sarro, and M.~Harman, ``Fairness improvement with multiple protected attributes: How far are we?'' in \emph{46th International Conference on Software Engineering (ICSE 2024)}.\hskip 1em plus 0.5em minus 0.4em\relax ACM, 2023.

\bibitem{chen2022maat}
Z.~Chen, J.~M. Zhang, F.~Sarro, and M.~Harman, ``Maat: a novel ensemble approach to addressing fairness and performance bugs for machine learning software,'' in \emph{Proceedings of the 30th ACM Joint European Software Engineering Conference and Symposium on the Foundations of Software Engineering}, 2022, pp. 1122--1134.

\bibitem{cooper2021emergent}
A.~F. Cooper, E.~Abrams, and N.~Na, ``Emergent unfairness in algorithmic fairness-accuracy trade-off research,'' in \emph{Proceedings of the 2021 AAAI/ACM Conference on AI, Ethics, and Society}, 2021, pp. 46--54.

\bibitem{liu2022accuracy}
S.~Liu and L.~N. Vicente, ``Accuracy and fairness trade-offs in machine learning: A stochastic multi-objective approach,'' \emph{Computational Management Science}, vol.~19, no.~3, pp. 513--537, 2022.

\bibitem{codebert}
\BIBentryALTinterwordspacing
Z.~Feng, D.~Guo, D.~Tang, N.~Duan, X.~Feng, M.~Gong, L.~Shou, B.~Qin, T.~Liu, D.~Jiang, and M.~Zhou, ``{C}ode{BERT}: A pre-trained model for programming and natural languages,'' in \emph{Findings of the Association for Computational Linguistics: EMNLP 2020}.\hskip 1em plus 0.5em minus 0.4em\relax Online: Association for Computational Linguistics, Nov. 2020, pp. 1536--1547. [Online]. Available: \url{https://aclanthology.org/2020.findings-emnlp.139}
\BIBentrySTDinterwordspacing

\bibitem{Ahmad2021UnifiedPF}
\BIBentryALTinterwordspacing
W.~U. Ahmad, S.~Chakraborty, B.~Ray, and K.-W. Chang, ``Unified pre-training for program understanding and generation,'' \emph{ArXiv}, vol. abs/2103.06333, 2021. [Online]. Available: \url{https://api.semanticscholar.org/CorpusID:232185260}
\BIBentrySTDinterwordspacing

\bibitem{CERT}
D.~Zan, B.~Chen, D.~Yang, Z.~Lin, M.~Kim, B.~Guan, Y.~Wang, W.~Chen, and J.-G. Lou, ``{CERT}: Continual pre-training on sketches for library-oriented code generation,'' in \emph{The 2022 International Joint Conference on Artificial Intelligence}, 2022.

\bibitem{bleu}
\BIBentryALTinterwordspacing
K.~Papineni, S.~Roukos, T.~Ward, and W.-J. Zhu, ``{B}leu: a method for automatic evaluation of machine translation,'' in \emph{Proceedings of the 40th Annual Meeting of the Association for Computational Linguistics}.\hskip 1em plus 0.5em minus 0.4em\relax Philadelphia, Pennsylvania, USA: Association for Computational Linguistics, Jul. 2002, pp. 311--318. [Online]. Available: \url{https://aclanthology.org/P02-1040}
\BIBentrySTDinterwordspacing

\bibitem{rouge}
\BIBentryALTinterwordspacing
C.-Y. Lin, ``{ROUGE}: A package for automatic evaluation of summaries,'' in \emph{Text Summarization Branches Out}.\hskip 1em plus 0.5em minus 0.4em\relax Barcelona, Spain: Association for Computational Linguistics, Jul. 2004, pp. 74--81. [Online]. Available: \url{https://aclanthology.org/W04-1013}
\BIBentrySTDinterwordspacing

\bibitem{Ren2020CodeBLEUAM}
\BIBentryALTinterwordspacing
S.~Ren, D.~Guo, S.~Lu, L.~Zhou, S.~Liu, D.~Tang, M.~Zhou, A.~Blanco, and S.~Ma, ``Codebleu: a method for automatic evaluation of code synthesis,'' \emph{ArXiv}, vol. abs/2009.10297, 2020. [Online]. Available: \url{https://api.semanticscholar.org/CorpusID:221836101}
\BIBentrySTDinterwordspacing

\bibitem{Evtikhiev2022OutOT}
\BIBentryALTinterwordspacing
M.~Evtikhiev, E.~Bogomolov, Y.~Sokolov, and T.~Bryksin, ``Out of the bleu: how should we assess quality of the code generation models?'' \emph{J. Syst. Softw.}, vol. 203, p. 111741, 2022. [Online]. Available: \url{https://api.semanticscholar.org/CorpusID:251371647}
\BIBentrySTDinterwordspacing

\bibitem{huang2024effibench}
D.~Huang, J.~M. Zhang, Y.~Qing, and H.~Cui, ``Effibench: Benchmarking the efficiency of automatically generated code,'' \emph{arXiv preprint arXiv:2402.02037}, 2024.

\end{thebibliography}


\begin{thebibliography}{100}
\providecommand{\url}[1]{#1}
\csname url@samestyle\endcsname
\providecommand{\newblock}{\relax}
\providecommand{\bibinfo}[2]{#2}
\providecommand{\BIBentrySTDinterwordspacing}{\spaceskip=0pt\relax}
\providecommand{\BIBentryALTinterwordstretchfactor}{4}
\providecommand{\BIBentryALTinterwordspacing}{\spaceskip=\fontdimen2\font plus
\BIBentryALTinterwordstretchfactor\fontdimen3\font minus \fontdimen4\font\relax}
\providecommand{\BIBforeignlanguage}[2]{{%
\expandafter\ifx\csname l@#1\endcsname\relax
\typeout{** WARNING: IEEEtran.bst: No hyphenation pattern has been}%
\typeout{** loaded for the language `#1'. Using the pattern for}%
\typeout{** the default language instead.}%
\else
\language=\csname l@#1\endcsname
\fi
#2}}
\providecommand{\BIBdecl}{\relax}
\BIBdecl

\bibitem{chen2021evaluating}
M.~Chen, J.~Tworek, H.~Jun, Q.~Yuan, H.~P. d.~O. Pinto, J.~Kaplan, H.~Edwards, Y.~Burda, N.~Joseph, G.~Brockman \emph{et~al.}, ``Evaluating large language models trained on code,'' \emph{arXiv preprint arXiv:2107.03374}, 2021.

\bibitem{OpenAI2023GPT4TR}
OpenAI, ``Gpt-4 technical report,'' \emph{ArXiv}, vol. abs/2303.08774, 2023.

\bibitem{huang2024effi}
D.~Huang, G.~Zeng, J.~Dai, M.~Luo, H.~Weng, Y.~Qing, H.~Cui, Z.~Guo, and J.~M. Zhang, ``Effi-code: Unleashing code efficiency in language models,'' \emph{arXiv preprint arXiv:2410.10209}, 2024.

\bibitem{huang2024effilearner}
D.~Huang, J.~Dai, H.~Weng, P.~Wu, Y.~Qing, H.~Cui, Z.~Guo, and J.~M. Zhang, ``Effilearner: Enhancing efficiency of generated code via self-optimization,'' \emph{arXiv preprint arXiv:2405.15189}, 2024.

\bibitem{huang2024rethinking}
D.~Huang, J.~M. Zhang, M.~Du, M.~Harman, and H.~Cui, ``Rethinking the influence of source code on test case generation,'' \emph{arXiv preprint arXiv:2409.09464}, 2024.

\bibitem{dai2024mhpp}
J.~Dai, J.~Lu, Y.~Feng, D.~Huang, G.~Zeng, R.~Ruan, M.~Cheng, H.~Tan, and Z.~Guo, ``Mhpp: Exploring the capabilities and limitations of language models beyond basic code generation,'' \emph{arXiv preprint arXiv:2405.11430}, 2024.

\bibitem{liu2023uncovering}
Y.~Liu, X.~Chen, Y.~Gao, Z.~Su, F.~Zhang, D.~Zan, J.-G. Lou, P.-Y. Chen, and T.-Y. Ho, ``Uncovering and quantifying social biases in code generation,'' vol.~36, pp. 2368--2380, 2023.

\bibitem{mukherjee2014role}
A.~N. Mukherjee, S.~Bhattacharyya, and R.~Bera, ``Role of information technology in human resource management of sme: A study on the use of applicant tracking system,'' \emph{IBMRD's Journal of Management \& Research}, pp. 1--22, 2014.

\bibitem{ahmad2009smart}
N.~Ahmad and A.~N. Abd~Alla, ``Smart evaluation for job vacancy application system,'' in \emph{2009 Second International Conference on the Applications of Digital Information and Web Technologies}.\hskip 1em plus 0.5em minus 0.4em\relax IEEE, 2009, pp. 452--455.

\bibitem{thakur2023language}
H.~Thakur, A.~Jain, P.~Vaddamanu, P.~P. Liang, and L.-P. Morency, ``Language models get a gender makeover: Mitigating gender bias with few-shot data interventions,'' \emph{arXiv preprint arXiv:2306.04597}, 2023.

\bibitem{ungless2022robust}
E.~L. Ungless, A.~Rafferty, H.~Nag, and B.~Ross, ``A robust bias mitigation procedure based on the stereotype content model,'' \emph{arXiv preprint arXiv:2210.14552}, 2022.

\bibitem{lee2023kosbi}
H.~Lee, S.~Hong, J.~Park, T.~Kim, G.~Kim, and J.-W. Ha, ``Kosbi: A dataset for mitigating social bias risks towards safer large language model application,'' \emph{arXiv preprint arXiv:2305.17701}, 2023.

\bibitem{barikeri2021redditbias}
S.~Barikeri, A.~Lauscher, I.~Vuli{\'c}, and G.~Glava{\v{s}}, ``Redditbias: A real-world resource for bias evaluation and debiasing of conversational language models,'' \emph{arXiv preprint arXiv:2106.03521}, 2021.

\bibitem{felkner2023winoqueer}
V.~K. Felkner, H.-C.~H. Chang, E.~Jang, and J.~May, ``Winoqueer: A community-in-the-loop benchmark for anti-lgbtq+ bias in large language models,'' \emph{arXiv preprint arXiv:2306.15087}, 2023.

\bibitem{fleisig2022mitigating}
E.~Fleisig and C.~Fellbaum, ``Mitigating gender bias in machine translation through adversarial learning,'' \emph{arXiv preprint arXiv:2203.10675}, 2022.

\bibitem{biswas23fairify}
S.~Biswas and H.~Rajan, ``Fairify: Fairness verification of neural networks,'' in \emph{ICSE'2023: The 45th International Conference on Software Engineering}, May 14-May 20 2023.

\bibitem{gohar23fairness}
S.~B. Usman~Gohar and H.~Rajan, ``Towards understanding fairness and its composition in ensemble machine learning,'' in \emph{ICSE'2023: The 45th International Conference on Software Engineering}, May 14-May 20 2023.

\bibitem{biswas20machine}
S.~Biswas and H.~Rajan, ``Do the machine learning models on a crowd sourced platform exhibit bias? an empirical study on model fairness,'' in \emph{ESEC/FSE'2020: The 28th ACM Joint European Software Engineering Conference and Symposium on the Foundations of Software Engineering}, November 8-November 13, 2020 2020.

\bibitem{2020t5}
\BIBentryALTinterwordspacing
C.~Raffel, N.~Shazeer, A.~Roberts, K.~Lee, S.~Narang, M.~Matena, Y.~Zhou, W.~Li, and P.~J. Liu, ``Exploring the limits of transfer learning with a unified text-to-text transformer,'' \emph{Journal of Machine Learning Research}, vol.~21, no. 140, pp. 1--67, 2020. [Online]. Available: \url{http://jmlr.org/papers/v21/20-074.html}
\BIBentrySTDinterwordspacing

\bibitem{wang2023codet5+}
Y.~Wang, H.~Le, A.~D. Gotmare, N.~D. Bui, J.~Li, and S.~C. Hoi, ``Codet5+: Open code large language models for code understanding and generation,'' \emph{arXiv preprint arXiv:2305.07922}, 2023.

\bibitem{ouyang2023llm}
S.~Ouyang, J.~M. Zhang, M.~Harman, and M.~Wang, ``Llm is like a box of chocolates: the non-determinism of chatgpt in code generation,'' \emph{arXiv preprint arXiv:2308.02828}, 2023.

\bibitem{Wang2022ReCodeRE}
\BIBentryALTinterwordspacing
S.~Wang, Z.~Li, H.~Qian, C.~Yang, Z.~Wang, M.~Shang, V.~Kumar, S.~Tan, B.~Ray, P.~Bhatia, R.~Nallapati, M.~K. Ramanathan, D.~Roth, and B.~Xiang, ``Recode: Robustness evaluation of code generation models,'' \emph{ArXiv}, vol. abs/2212.10264, 2022. [Online]. Available: \url{https://api.semanticscholar.org/CorpusID:254877229}
\BIBentrySTDinterwordspacing

\bibitem{alayrac2022flamingo}
J.-B. Alayrac, J.~Donahue, P.~Luc, A.~Miech, I.~Barr, Y.~Hasson, K.~Lenc, A.~Mensch, K.~Millican, M.~Reynolds \emph{et~al.}, ``Flamingo: a visual language model for few-shot learning,'' \emph{Advances in Neural Information Processing Systems}, vol.~35, pp. 23\,716--23\,736, 2022.

\bibitem{izacard2022few}
G.~Izacard, P.~Lewis, M.~Lomeli, L.~Hosseini, F.~Petroni, T.~Schick, J.~Dwivedi-Yu, A.~Joulin, S.~Riedel, and E.~Grave, ``Few-shot learning with retrieval augmented language models,'' \emph{arXiv preprint arXiv:2208.03299}, 2022.

\bibitem{tunstall2022efficient}
L.~Tunstall, N.~Reimers, U.~E.~S. Jo, L.~Bates, D.~Korat, M.~Wasserblat, and O.~Pereg, ``Efficient few-shot learning without prompts,'' \emph{arXiv preprint arXiv:2209.11055}, 2022.

\bibitem{wei2022chain}
J.~Wei, X.~Wang, D.~Schuurmans, M.~Bosma, F.~Xia, E.~Chi, Q.~V. Le, D.~Zhou \emph{et~al.}, ``Chain-of-thought prompting elicits reasoning in large language models,'' \emph{Advances in Neural Information Processing Systems}, vol.~35, pp. 24\,824--24\,837, 2022.

\bibitem{madaan2022text}
A.~Madaan and A.~Yazdanbakhsh, ``Text and patterns: For effective chain of thought, it takes two to tango,'' \emph{arXiv preprint arXiv:2209.07686}, 2022.

\bibitem{wang2022self}
X.~Wang, J.~Wei, D.~Schuurmans, Q.~Le, E.~Chi, S.~Narang, A.~Chowdhery, and D.~Zhou, ``Self-consistency improves chain of thought reasoning in language models,'' \emph{arXiv preprint arXiv:2203.11171}, 2022.

\bibitem{chu2023survey}
Z.~Chu, J.~Chen, Q.~Chen, W.~Yu, T.~He, H.~Wang, W.~Peng, M.~Liu, B.~Qin, and T.~Liu, ``A survey of chain of thought reasoning: Advances, frontiers and future,'' \emph{arXiv preprint arXiv:2309.15402}, 2023.

\bibitem{huang2023codecot}
D.~Huang, Q.~Bu, and H.~Cui, ``Codecot and beyond: Learning to program and test like a developer,'' \emph{arXiv preprint arXiv:2308.08784}, 2023.

\bibitem{huang2023agentcoder}
D.~Huang, Q.~Bu, J.~M. Zhang, M.~Luck, and H.~Cui, ``Agentcoder: Multi-agent-based code generation with iterative testing and optimisation,'' \emph{arXiv preprint arXiv:2312.13010}, 2023.

\bibitem{Lialphacode2022}
\BIBentryALTinterwordspacing
Y.~Li, D.~H. Choi, J.~Chung, N.~Kushman, J.~Schrittwieser, R.~Leblond, T.~Eccles, J.~Keeling, F.~Gimeno, A.~D. Lago, T.~Hubert, P.~Choy, C.~de~Masson~d'Autume, I.~Babuschkin, X.~Chen, P.~Huang, J.~Welbl, S.~Gowal, A.~Cherepanov, J.~Molloy, D.~J. Mankowitz, E.~S. Robson, P.~Kohli, N.~de~Freitas, K.~Kavukcuoglu, and O.~Vinyals, ``Competition-level code generation with alphacode,'' \emph{CoRR}, vol. abs/2203.07814, 2022. [Online]. Available: \url{https://doi.org/10.48550/arXiv.2203.07814}
\BIBentrySTDinterwordspacing

\bibitem{nijkamp2022codegen}
E.~Nijkamp, B.~Pang, H.~Hayashi, L.~Tu, H.~Wang, Y.~Zhou, S.~Savarese, and C.~Xiong, ``Codegen: An open large language model for code with multi-turn program synthesis,'' \emph{ICLR}, 2023.

\bibitem{FriedAL0WSZYZL23}
\BIBentryALTinterwordspacing
D.~Fried, A.~Aghajanyan, J.~Lin, S.~Wang, E.~Wallace, F.~Shi, R.~Zhong, S.~Yih, L.~Zettlemoyer, and M.~Lewis, ``Incoder: {A} generative model for code infilling and synthesis,'' in \emph{The Eleventh International Conference on Learning Representations, {ICLR} 2023, Kigali, Rwanda, May 1-5, 2023}.\hskip 1em plus 0.5em minus 0.4em\relax OpenReview.net, 2023. [Online]. Available: \url{https://openreview.net/pdf?id=hQwb-lbM6EL}
\BIBentrySTDinterwordspacing

\bibitem{LiStarCoder2023}
\BIBentryALTinterwordspacing
R.~Li, L.~B. Allal, Y.~Zi, N.~Muennighoff, D.~Kocetkov, C.~Mou, M.~Marone, C.~Akiki, J.~Li, J.~Chim, Q.~Liu, E.~Zheltonozhskii, T.~Y. Zhuo, T.~Wang, O.~Dehaene, M.~Davaadorj, J.~Lamy{-}Poirier, J.~Monteiro, O.~Shliazhko, N.~Gontier, N.~Meade, A.~Zebaze, M.~Yee, L.~K. Umapathi, J.~Zhu, B.~Lipkin, M.~Oblokulov, Z.~Wang, R.~M. V, J.~Stillerman, S.~S. Patel, D.~Abulkhanov, M.~Zocca, M.~Dey, Z.~Zhang, N.~Moustafa{-}Fahmy, U.~Bhattacharyya, W.~Yu, S.~Singh, S.~Luccioni, P.~Villegas, M.~Kunakov, F.~Zhdanov, M.~Romero, T.~Lee, N.~Timor, J.~Ding, C.~Schlesinger, H.~Schoelkopf, J.~Ebert, T.~Dao, M.~Mishra, A.~Gu, J.~Robinson, C.~J. Anderson, B.~Dolan{-}Gavitt, D.~Contractor, S.~Reddy, D.~Fried, D.~Bahdanau, Y.~Jernite, C.~M. Ferrandis, S.~Hughes, T.~Wolf, A.~Guha, L.~von Werra, and H.~de~Vries, ``Starcoder: may the source be with you!'' \emph{CoRR}, vol. abs/2305.06161, 2023. [Online]. Available: \url{https://doi.org/10.48550/arXiv.2305.06161}
\BIBentrySTDinterwordspacing

\bibitem{Loubnasanta2023}
\BIBentryALTinterwordspacing
L.~B. Allal, R.~Li, D.~Kocetkov, C.~Mou, C.~Akiki, C.~M. Ferrandis, N.~Muennighoff, M.~Mishra, A.~Gu, M.~Dey, L.~K. Umapathi, C.~J. Anderson, Y.~Zi, J.~Lamy{-}Poirier, H.~Schoelkopf, S.~Troshin, D.~Abulkhanov, M.~Romero, M.~Lappert, F.~D. Toni, B.~G. del R{\'{\i}}o, Q.~Liu, S.~Bose, U.~Bhattacharyya, T.~Y. Zhuo, I.~Yu, P.~Villegas, M.~Zocca, S.~Mangrulkar, D.~Lansky, H.~Nguyen, D.~Contractor, L.~Villa, J.~Li, D.~Bahdanau, Y.~Jernite, S.~Hughes, D.~Fried, A.~Guha, H.~de~Vries, and L.~von Werra, ``Santacoder: don't reach for the stars!'' \emph{CoRR}, vol. abs/2301.03988, 2023. [Online]. Available: \url{https://doi.org/10.48550/arXiv.2301.03988}
\BIBentrySTDinterwordspacing

\bibitem{deepseekcoder}
\BIBentryALTinterwordspacing
DeepSeekAI, ``Deepseek coder: Let the code write itself,'' 2023. [Online]. Available: \url{https://deepseekcoder.github.io/}
\BIBentrySTDinterwordspacing

\bibitem{Rozire2023CodeLO}
\BIBentryALTinterwordspacing
B.~Rozi{\`e}re, J.~Gehring, F.~Gloeckle, S.~Sootla, I.~Gat, X.~Tan, Y.~Adi, J.~Liu, T.~Remez, J.~Rapin, A.~Kozhevnikov, I.~Evtimov, J.~Bitton, M.~P. Bhatt, C.~C. Ferrer, A.~Grattafiori, W.~Xiong, A.~D'efossez, J.~Copet, F.~Azhar, H.~Touvron, L.~Martin, N.~Usunier, T.~Scialom, and G.~Synnaeve, ``Code llama: Open foundation models for code,'' \emph{ArXiv}, vol. abs/2308.12950, 2023. [Online]. Available: \url{https://api.semanticscholar.org/CorpusID:261100919}
\BIBentrySTDinterwordspacing

\bibitem{BrownMRSKDNSSAA20}
\BIBentryALTinterwordspacing
T.~B. Brown, B.~Mann, N.~Ryder, M.~Subbiah, J.~Kaplan, P.~Dhariwal, A.~Neelakantan, P.~Shyam, G.~Sastry, A.~Askell, S.~Agarwal, A.~Herbert{-}Voss, G.~Krueger, T.~Henighan, R.~Child, A.~Ramesh, D.~M. Ziegler, J.~Wu, C.~Winter, C.~Hesse, M.~Chen, E.~Sigler, M.~Litwin, S.~Gray, B.~Chess, J.~Clark, C.~Berner, S.~McCandlish, A.~Radford, I.~Sutskever, and D.~Amodei, ``Language models are few-shot learners,'' in \emph{Advances in Neural Information Processing Systems 33: Annual Conference on Neural Information Processing Systems 2020, NeurIPS 2020, December 6-12, 2020, virtual}, H.~Larochelle, M.~Ranzato, R.~Hadsell, M.~Balcan, and H.~Lin, Eds., 2020. [Online]. Available: \url{https://proceedings.neurips.cc/paper/2020/hash/1457c0d6bfcb4967418bfb8ac142f64a-Abstract.html}
\BIBentrySTDinterwordspacing

\bibitem{Touvron2023}
\BIBentryALTinterwordspacing
H.~Touvron, L.~Martin, K.~Stone, P.~Albert, A.~Almahairi, Y.~Babaei, N.~Bashlykov, S.~Batra, P.~Bhargava, S.~Bhosale, D.~Bikel, L.~Blecher, C.~Canton{-}Ferrer, M.~Chen, G.~Cucurull, D.~Esiobu, J.~Fernandes, J.~Fu, W.~Fu, B.~Fuller, C.~Gao, V.~Goswami, N.~Goyal, A.~Hartshorn, S.~Hosseini, R.~Hou, H.~Inan, M.~Kardas, V.~Kerkez, M.~Khabsa, I.~Kloumann, A.~Korenev, P.~S. Koura, M.~Lachaux, T.~Lavril, J.~Lee, D.~Liskovich, Y.~Lu, Y.~Mao, X.~Martinet, T.~Mihaylov, P.~Mishra, I.~Molybog, Y.~Nie, A.~Poulton, J.~Reizenstein, R.~Rungta, K.~Saladi, A.~Schelten, R.~Silva, E.~M. Smith, R.~Subramanian, X.~E. Tan, B.~Tang, R.~Taylor, A.~Williams, J.~X. Kuan, P.~Xu, Z.~Yan, I.~Zarov, Y.~Zhang, A.~Fan, M.~Kambadur, S.~Narang, A.~Rodriguez, R.~Stojnic, S.~Edunov, and T.~Scialom, ``Llama 2: Open foundation and fine-tuned chat models,'' \emph{CoRR}, vol. abs/2307.09288, 2023. [Online]. Available: \url{https://doi.org/10.48550/arXiv.2307.09288}
\BIBentrySTDinterwordspacing

\bibitem{GPT35turbo}
\BIBentryALTinterwordspacing
OpenAI, ``{GPT-3.5} {T}urbo,'' 2023. [Online]. Available: \url{https://platform.openai.com/docs/models/gpt-3-5}
\BIBentrySTDinterwordspacing

\bibitem{GPT4}
\BIBentryALTinterwordspacing
------, ``{GPT-4} {T}echnical {R}eport,'' \emph{CoRR}, vol. abs/2303.08774, 2023. [Online]. Available: \url{https://doi.org/10.48550/arXiv.2303.08774}
\BIBentrySTDinterwordspacing

\bibitem{Luo2023WizardCoderEC}
\BIBentryALTinterwordspacing
Z.~Luo, C.~Xu, P.~Zhao, Q.~Sun, X.~Geng, W.~Hu, C.~Tao, J.~Ma, Q.~Lin, and D.~Jiang, ``Wizardcoder: Empowering code large language models with evol-instruct,'' \emph{ArXiv}, vol. abs/2306.08568, 2023. [Online]. Available: \url{https://api.semanticscholar.org/CorpusID:259164815}
\BIBentrySTDinterwordspacing

\bibitem{Phi3}
\BIBentryALTinterwordspacing
M.~I. Abdin, S.~A. Jacobs, A.~A. Awan, J.~Aneja, A.~Awadallah, H.~Awadalla, N.~Bach, A.~Bahree, A.~Bakhtiari, H.~S. Behl, A.~Benhaim, M.~Bilenko, J.~Bjorck, S.~Bubeck, M.~Cai, C.~C.~T. Mendes, W.~Chen, V.~Chaudhary, P.~Chopra, A.~D. Giorno, G.~de~Rosa, M.~Dixon, R.~Eldan, D.~Iter, A.~Garg, A.~Goswami, S.~Gunasekar, E.~Haider, J.~Hao, R.~J. Hewett, J.~Huynh, M.~Javaheripi, X.~Jin, P.~Kauffmann, N.~Karampatziakis, D.~Kim, M.~Khademi, L.~Kurilenko, J.~R. Lee, Y.~T. Lee, Y.~Li, C.~Liang, W.~Liu, E.~Lin, Z.~Lin, P.~Madan, A.~Mitra, H.~Modi, A.~Nguyen, B.~Norick, B.~Patra, D.~Perez{-}Becker, T.~Portet, R.~Pryzant, H.~Qin, M.~Radmilac, C.~Rosset, S.~Roy, O.~Ruwase, O.~Saarikivi, A.~Saied, A.~Salim, M.~Santacroce, S.~Shah, N.~Shang, H.~Sharma, X.~Song, M.~Tanaka, X.~Wang, R.~Ward, G.~Wang, P.~Witte, M.~Wyatt, C.~Xu, J.~Xu, S.~Yadav, F.~Yang, Z.~Yang, D.~Yu, C.~Zhang, C.~Zhang, J.~Zhang, L.~L. Zhang, Y.~Zhang, Y.~Zhang, Y.~Zhang, and X.~Zhou, ``Phi-3 technical report: {A} highly capable language model locally on your
  phone,'' \emph{CoRR}, vol. abs/2404.14219, 2024. [Online]. Available: \url{https://doi.org/10.48550/arXiv.2404.14219}
\BIBentrySTDinterwordspacing

\bibitem{Haque2022}
\BIBentryALTinterwordspacing
M.~M.~A. Haque, W.~U. Ahmad, I.~Lourentzou, and C.~Brown, ``Fixeval: Execution-based evaluation of program fixes for competitive programming problems,'' \emph{CoRR}, vol. abs/2206.07796, 2022. [Online]. Available: \url{https://doi.org/10.48550/arXiv.2206.07796}
\BIBentrySTDinterwordspacing

\bibitem{JiangLLT23}
\BIBentryALTinterwordspacing
N.~Jiang, K.~Liu, T.~Lutellier, and L.~Tan, ``Impact of code language models on automated program repair,'' in \emph{45th {IEEE/ACM} International Conference on Software Engineering, {ICSE} 2023, Melbourne, Australia, May 14-20, 2023}.\hskip 1em plus 0.5em minus 0.4em\relax {IEEE}, 2023, pp. 1430--1442. [Online]. Available: \url{https://doi.org/10.1109/ICSE48619.2023.00125}
\BIBentrySTDinterwordspacing

\bibitem{LemieuxILS23}
\BIBentryALTinterwordspacing
C.~Lemieux, J.~P. Inala, S.~K. Lahiri, and S.~Sen, ``Codamosa: Escaping coverage plateaus in test generation with pre-trained large language models,'' in \emph{45th {IEEE/ACM} International Conference on Software Engineering, {ICSE} 2023, Melbourne, Australia, May 14-20, 2023}.\hskip 1em plus 0.5em minus 0.4em\relax {IEEE}, 2023, pp. 919--931. [Online]. Available: \url{https://doi.org/10.1109/ICSE48619.2023.00085}
\BIBentrySTDinterwordspacing

\bibitem{Deng2023}
\BIBentryALTinterwordspacing
Y.~Deng, C.~S. Xia, C.~Yang, S.~D. Zhang, S.~Yang, and L.~Zhang, ``Large language models are edge-case fuzzers: Testing deep learning libraries via fuzzgpt,'' \emph{CoRR}, vol. abs/2304.02014, 2023. [Online]. Available: \url{https://doi.org/10.48550/arXiv.2304.02014}
\BIBentrySTDinterwordspacing

\bibitem{RoziereLCL20}
\BIBentryALTinterwordspacing
B.~Rozi{\`{e}}re, M.~Lachaux, L.~Chanussot, and G.~Lample, ``Unsupervised translation of programming languages,'' in \emph{Advances in Neural Information Processing Systems 33: Annual Conference on Neural Information Processing Systems 2020, NeurIPS 2020, December 6-12, 2020, virtual}, H.~Larochelle, M.~Ranzato, R.~Hadsell, M.~Balcan, and H.~Lin, Eds., 2020. [Online]. Available: \url{https://proceedings.neurips.cc/paper/2020/hash/ed23fbf18c2cd35f8c7f8de44f85c08d-Abstract.html}
\BIBentrySTDinterwordspacing

\bibitem{AhmadTCC23}
\BIBentryALTinterwordspacing
W.~U. Ahmad, M.~G.~R. Tushar, S.~Chakraborty, and K.~Chang, ``{AVATAR:} {A} parallel corpus for java-python program translation,'' in \emph{Findings of the Association for Computational Linguistics: {ACL} 2023, Toronto, Canada, July 9-14, 2023}, A.~Rogers, J.~L. Boyd{-}Graber, and N.~Okazaki, Eds.\hskip 1em plus 0.5em minus 0.4em\relax Association for Computational Linguistics, 2023, pp. 2268--2281. [Online]. Available: \url{https://doi.org/10.18653/v1/2023.findings-acl.143}
\BIBentrySTDinterwordspacing

\bibitem{MirLPG22}
\BIBentryALTinterwordspacing
A.~M. Mir, E.~Latoskinas, S.~Proksch, and G.~Gousios, ``Type4py: Practical deep similarity learning-based type inference for python,'' in \emph{44th {IEEE/ACM} 44th International Conference on Software Engineering, {ICSE} 2022, Pittsburgh, PA, USA, May 25-27, 2022}.\hskip 1em plus 0.5em minus 0.4em\relax {ACM}, 2022, pp. 2241--2252. [Online]. Available: \url{https://doi.org/10.1145/3510003.3510124}
\BIBentrySTDinterwordspacing

\bibitem{WeiDD23}
\BIBentryALTinterwordspacing
J.~Wei, G.~Durrett, and I.~Dillig, ``Typet5: Seq2seq type inference using static analysis,'' in \emph{The Eleventh International Conference on Learning Representations, {ICLR} 2023, Kigali, Rwanda, May 1-5, 2023}.\hskip 1em plus 0.5em minus 0.4em\relax OpenReview.net, 2023. [Online]. Available: \url{https://openreview.net/pdf?id=4TyNEhI2GdN}
\BIBentrySTDinterwordspacing

\bibitem{HasanMIMHHAIS21}
\BIBentryALTinterwordspacing
M.~Hasan, T.~Muttaqueen, A.~A. Ishtiaq, K.~S. Mehrab, M.~M.~A. Haque, T.~Hasan, W.~U. Ahmad, A.~Iqbal, and R.~Shahriyar, ``Codesc: {A} large code-description parallel dataset,'' in \emph{Findings of the Association for Computational Linguistics: {ACL/IJCNLP} 2021, Online Event, August 1-6, 2021}, ser. Findings of {ACL}, C.~Zong, F.~Xia, W.~Li, and R.~Navigli, Eds., vol. {ACL/IJCNLP} 2021.\hskip 1em plus 0.5em minus 0.4em\relax Association for Computational Linguistics, 2021, pp. 210--218. [Online]. Available: \url{https://doi.org/10.18653/v1/2021.findings-acl.18}
\BIBentrySTDinterwordspacing

\bibitem{AhmedD22}
\BIBentryALTinterwordspacing
T.~Ahmed and P.~T. Devanbu, ``Few-shot training llms for project-specific code-summarization,'' in \emph{37th {IEEE/ACM} International Conference on Automated Software Engineering, {ASE} 2022, Rochester, MI, USA, October 10-14, 2022}.\hskip 1em plus 0.5em minus 0.4em\relax {ACM}, 2022, pp. 177:1--177:5. [Online]. Available: \url{https://doi.org/10.1145/3551349.3559555}
\BIBentrySTDinterwordspacing

\bibitem{Austin2021ProgramSW}
\BIBentryALTinterwordspacing
J.~Austin, A.~Odena, M.~Nye, M.~Bosma, H.~Michalewski, D.~Dohan, E.~Jiang, C.~J. Cai, M.~Terry, Q.~V. Le, and C.~Sutton, ``Program synthesis with large language models,'' \emph{ArXiv}, vol. abs/2108.07732, 2021. [Online]. Available: \url{https://api.semanticscholar.org/CorpusID:237142385}
\BIBentrySTDinterwordspacing

\bibitem{liu2023is}
\BIBentryALTinterwordspacing
J.~Liu, C.~S. Xia, Y.~Wang, and L.~ZHANG, ``Is your code generated by chat{GPT} really correct? rigorous evaluation of large language models for code generation,'' in \emph{Thirty-seventh Conference on Neural Information Processing Systems}, 2023. [Online]. Available: \url{https://openreview.net/forum?id=1qvx610Cu7}
\BIBentrySTDinterwordspacing

\bibitem{WangRecode2023}
\BIBentryALTinterwordspacing
S.~Wang, Z.~Li, H.~Qian, C.~Yang, Z.~Wang, M.~Shang, V.~Kumar, S.~Tan, B.~Ray, P.~Bhatia, R.~Nallapati, M.~K. Ramanathan, D.~Roth, and B.~Xiang, ``Recode: Robustness evaluation of code generation models,'' in \emph{Proceedings of the 61st Annual Meeting of the Association for Computational Linguistics (Volume 1: Long Papers), {ACL} 2023, Toronto, Canada, July 9-14, 2023}, A.~Rogers, J.~L. Boyd{-}Graber, and N.~Okazaki, Eds.\hskip 1em plus 0.5em minus 0.4em\relax Association for Computational Linguistics, 2023, pp. 13\,818--13\,843. [Online]. Available: \url{https://doi.org/10.18653/v1/2023.acl-long.773}
\BIBentrySTDinterwordspacing

\bibitem{ZhengHEX2023}
\BIBentryALTinterwordspacing
Q.~Zheng, X.~Xia, X.~Zou, Y.~Dong, S.~Wang, Y.~Xue, Z.~Wang, L.~Shen, A.~Wang, Y.~Li, T.~Su, Z.~Yang, and J.~Tang, ``Codegeex: {A} pre-trained model for code generation with multilingual evaluations on humaneval-x,'' \emph{CoRR}, vol. abs/2303.17568, 2023. [Online]. Available: \url{https://doi.org/10.48550/arXiv.2303.17568}
\BIBentrySTDinterwordspacing

\bibitem{CassanoGNNPPYZAFGGJ23}
\BIBentryALTinterwordspacing
F.~Cassano, J.~Gouwar, D.~Nguyen, S.~Nguyen, L.~Phipps{-}Costin, D.~Pinckney, M.~Yee, Y.~Zi, C.~J. Anderson, M.~Q. Feldman, A.~Guha, M.~Greenberg, and A.~Jangda, ``Multipl-e: {A} scalable and polyglot approach to benchmarking neural code generation,'' \emph{{IEEE} Trans. Software Eng.}, vol.~49, no.~7, pp. 3675--3691, 2023. [Online]. Available: \url{https://doi.org/10.1109/TSE.2023.3267446}
\BIBentrySTDinterwordspacing

\bibitem{AthiwaratkunGWL23}
\BIBentryALTinterwordspacing
B.~Athiwaratkun, S.~K. Gouda, Z.~Wang, X.~Li, Y.~Tian, M.~Tan, W.~U. Ahmad, S.~Wang, Q.~Sun, M.~Shang, S.~K. Gonugondla, H.~Ding, V.~Kumar, N.~Fulton, A.~Farahani, S.~Jain, R.~Giaquinto, H.~Qian, M.~K. Ramanathan, and R.~Nallapati, ``Multi-lingual evaluation of code generation models,'' in \emph{The Eleventh International Conference on Learning Representations, {ICLR} 2023, Kigali, Rwanda, May 1-5, 2023}.\hskip 1em plus 0.5em minus 0.4em\relax OpenReview.net, 2023. [Online]. Available: \url{https://openreview.net/pdf?id=Bo7eeXm6An8}
\BIBentrySTDinterwordspacing

\bibitem{Lai0WZZZYFWY23}
\BIBentryALTinterwordspacing
Y.~Lai, C.~Li, Y.~Wang, T.~Zhang, R.~Zhong, L.~Zettlemoyer, W.~Yih, D.~Fried, S.~I. Wang, and T.~Yu, ``{DS-1000:} {A} natural and reliable benchmark for data science code generation,'' in \emph{International Conference on Machine Learning, {ICML} 2023, 23-29 July 2023, Honolulu, Hawaii, {USA}}, ser. Proceedings of Machine Learning Research, A.~Krause, E.~Brunskill, K.~Cho, B.~Engelhardt, S.~Sabato, and J.~Scarlett, Eds., vol. 202.\hskip 1em plus 0.5em minus 0.4em\relax {PMLR}, 2023, pp. 18\,319--18\,345. [Online]. Available: \url{https://proceedings.mlr.press/v202/lai23b.html}
\BIBentrySTDinterwordspacing

\bibitem{YinLXRWSHBCMPS23}
\BIBentryALTinterwordspacing
P.~Yin, W.~Li, K.~Xiao, A.~Rao, Y.~Wen, K.~Shi, J.~Howland, P.~Bailey, M.~Catasta, H.~Michalewski, O.~Polozov, and C.~Sutton, ``Natural language to code generation in interactive data science notebooks,'' in \emph{Proceedings of the 61st Annual Meeting of the Association for Computational Linguistics (Volume 1: Long Papers), {ACL} 2023, Toronto, Canada, July 9-14, 2023}, A.~Rogers, J.~L. Boyd{-}Graber, and N.~Okazaki, Eds.\hskip 1em plus 0.5em minus 0.4em\relax Association for Computational Linguistics, 2023, pp. 126--173. [Online]. Available: \url{https://doi.org/10.18653/v1/2023.acl-long.9}
\BIBentrySTDinterwordspacing

\bibitem{ZanCYLKGWCL22}
\BIBentryALTinterwordspacing
D.~Zan, B.~Chen, D.~Yang, Z.~Lin, M.~Kim, B.~Guan, Y.~Wang, W.~Chen, and J.~Lou, ``{CERT:} continual pre-training on sketches for library-oriented code generation,'' in \emph{Proceedings of the Thirty-First International Joint Conference on Artificial Intelligence, {IJCAI} 2022, Vienna, Austria, 23-29 July 2022}, L.~D. Raedt, Ed.\hskip 1em plus 0.5em minus 0.4em\relax ijcai.org, 2022, pp. 2369--2375. [Online]. Available: \url{https://doi.org/10.24963/ijcai.2022/329}
\BIBentrySTDinterwordspacing

\bibitem{JainVINPR022}
\BIBentryALTinterwordspacing
N.~Jain, S.~Vaidyanath, A.~S. Iyer, N.~Natarajan, S.~Parthasarathy, S.~K. Rajamani, and R.~Sharma, ``Jigsaw: Large language models meet program synthesis,'' in \emph{44th {IEEE/ACM} 44th International Conference on Software Engineering, {ICSE} 2022, Pittsburgh, PA, USA, May 25-27, 2022}.\hskip 1em plus 0.5em minus 0.4em\relax {ACM}, 2022, pp. 1219--1231. [Online]. Available: \url{https://doi.org/10.1145/3510003.3510203}
\BIBentrySTDinterwordspacing

\bibitem{Patil2023}
\BIBentryALTinterwordspacing
S.~G. Patil, T.~Zhang, X.~Wang, and J.~E. Gonzalez, ``Gorilla: Large language model connected with massive apis,'' \emph{CoRR}, vol. abs/2305.15334, 2023. [Online]. Available: \url{https://doi.org/10.48550/arXiv.2305.15334}
\BIBentrySTDinterwordspacing

\bibitem{NijkampPHTWZSX23}
\BIBentryALTinterwordspacing
E.~Nijkamp, B.~Pang, H.~Hayashi, L.~Tu, H.~Wang, Y.~Zhou, S.~Savarese, and C.~Xiong, ``Codegen: An open large language model for code with multi-turn program synthesis,'' in \emph{The Eleventh International Conference on Learning Representations, {ICLR} 2023, Kigali, Rwanda, May 1-5, 2023}.\hskip 1em plus 0.5em minus 0.4em\relax OpenReview.net, 2023. [Online]. Available: \url{https://openreview.net/pdf?id=iaYcJKpY2B\_}
\BIBentrySTDinterwordspacing

\bibitem{LiuRepo2023}
\BIBentryALTinterwordspacing
T.~Liu, C.~Xu, and J.~J. McAuley, ``Repobench: Benchmarking repository-level code auto-completion systems,'' \emph{CoRR}, vol. abs/2306.03091, 2023. [Online]. Available: \url{https://doi.org/10.48550/arXiv.2306.03091}
\BIBentrySTDinterwordspacing

\bibitem{Jimenez2023}
\BIBentryALTinterwordspacing
C.~E. Jimenez, J.~Yang, A.~Wettig, S.~Yao, K.~Pei, O.~Press, and K.~Narasimhan, ``Swe-bench: Can language models resolve real-world github issues?'' \emph{CoRR}, vol. abs/2310.06770, 2023. [Online]. Available: \url{https://doi.org/10.48550/arXiv.2310.06770}
\BIBentrySTDinterwordspacing

\bibitem{ShrivastavaLT23}
\BIBentryALTinterwordspacing
D.~Shrivastava, H.~Larochelle, and D.~Tarlow, ``Repository-level prompt generation for large language models of code,'' in \emph{International Conference on Machine Learning, {ICML} 2023, 23-29 July 2023, Honolulu, Hawaii, {USA}}, ser. Proceedings of Machine Learning Research, A.~Krause, E.~Brunskill, K.~Cho, B.~Engelhardt, S.~Sabato, and J.~Scarlett, Eds., vol. 202.\hskip 1em plus 0.5em minus 0.4em\relax {PMLR}, 2023, pp. 31\,693--31\,715. [Online]. Available: \url{https://proceedings.mlr.press/v202/shrivastava23a.html}
\BIBentrySTDinterwordspacing

\bibitem{ZhangCZKLZMLC23}
\BIBentryALTinterwordspacing
F.~Zhang, B.~Chen, Y.~Zhang, J.~Keung, J.~Liu, D.~Zan, Y.~Mao, J.~Lou, and W.~Chen, ``Repocoder: Repository-level code completion through iterative retrieval and generation,'' in \emph{Proceedings of the 2023 Conference on Empirical Methods in Natural Language Processing, {EMNLP} 2023, Singapore, December 6-10, 2023}, H.~Bouamor, J.~Pino, and K.~Bali, Eds.\hskip 1em plus 0.5em minus 0.4em\relax Association for Computational Linguistics, 2023, pp. 2471--2484. [Online]. Available: \url{https://aclanthology.org/2023.emnlp-main.151}
\BIBentrySTDinterwordspacing

\bibitem{Ding2022}
\BIBentryALTinterwordspacing
Y.~Ding, Z.~Wang, W.~U. Ahmad, M.~K. Ramanathan, R.~Nallapati, P.~Bhatia, D.~Roth, and B.~Xiang, ``Cocomic: Code completion by jointly modeling in-file and cross-file context,'' \emph{CoRR}, vol. abs/2212.10007, 2022. [Online]. Available: \url{https://doi.org/10.48550/arXiv.2212.10007}
\BIBentrySTDinterwordspacing

\bibitem{obrien24prompt}
D.~OBrien, S.~Biswas, S.~Imtiaz, R.~Abdalkareem, E.~Shihab, and H.~Rajan, ``Are prompt engineering and todo comments friends or foes? an evaluation on github copilot,'' in \emph{ICSE'2024: The 46th International Conference on Software Engineering}, April 14-April 20 2024.

\bibitem{chen2023comprehensive}
Z.~Chen, J.~M. Zhang, F.~Sarro, and M.~Harman, ``A comprehensive empirical study of bias mitigation methods for machine learning classifiers,'' \emph{ACM Transactions on Software Engineering and Methodology}, vol.~32, no.~4, pp. 1--30, 2023.

\bibitem{ding2021retiring}
F.~Ding, M.~Hardt, J.~Miller, and L.~Schmidt, ``Retiring adult: New datasets for fair machine learning,'' \emph{Advances in neural information processing systems}, vol.~34, pp. 6478--6490, 2021.

\bibitem{le2022survey}
T.~Le~Quy, A.~Roy, V.~Iosifidis, W.~Zhang, and E.~Ntoutsi, ``A survey on datasets for fairness-aware machine learning,'' \emph{Wiley Interdisciplinary Reviews: Data Mining and Knowledge Discovery}, vol.~12, no.~3, p. e1452, 2022.

\bibitem{friedler2019comparative}
S.~A. Friedler, C.~Scheidegger, S.~Venkatasubramanian, S.~Choudhary, E.~P. Hamilton, and D.~Roth, ``A comparative study of fairness-enhancing interventions in machine learning,'' in \emph{Proceedings of the conference on fairness, accountability, and transparency}, 2019, pp. 329--338.

\bibitem{besse2022survey}
P.~Besse, E.~del Barrio, P.~Gordaliza, J.-M. Loubes, and L.~Risser, ``A survey of bias in machine learning through the prism of statistical parity,'' \emph{The American Statistician}, vol.~76, no.~2, pp. 188--198, 2022.

\bibitem{kang2021multifair}
J.~Kang, T.~Xie, X.~Wu, R.~Maciejewski, and H.~Tong, ``Multifair: Multi-group fairness in machine learning,'' \emph{arXiv preprint arXiv:2105.11069}, 2021.

\bibitem{mehrabi2021survey}
N.~Mehrabi, F.~Morstatter, N.~Saxena, K.~Lerman, and A.~Galstyan, ``A survey on bias and fairness in machine learning,'' \emph{ACM computing surveys (CSUR)}, vol.~54, no.~6, pp. 1--35, 2021.

\bibitem{kearns2019empirical}
M.~Kearns, S.~Neel, A.~Roth, and Z.~S. Wu, ``An empirical study of rich subgroup fairness for machine learning,'' in \emph{Proceedings of the conference on fairness, accountability, and transparency}, 2019, pp. 100--109.

\bibitem{komiyama2017two}
J.~Komiyama and H.~Shimao, ``Two-stage algorithm for fairness-aware machine learning,'' \emph{arXiv preprint arXiv:1710.04924}, 2017.

\bibitem{xia2022summer}
F.~Xia, T.~Guo, X.~Bai, A.~Shatte, Z.~Liu, and J.~Tang, ``Summer: Bias-aware prediction of graduate employment based on educational big data,'' \emph{ACM/IMS Transactions on Data Science (TDS)}, vol.~2, no.~4, pp. 1--24, 2022.

\bibitem{papadaki2022federated}
A.~Papadaki, N.~Martinez, M.~A. Bertran, G.~Sapiro, and M.~R. Rodrigues, ``Federated fairness without access to demographics,'' in \emph{Workshop on Federated Learning: Recent Advances and New Challenges (in Conjunction with NeurIPS 2022)}, 2022.

\bibitem{han2023retiring}
X.~Han, Z.~Jiang, H.~Jin, Z.~Liu, N.~Zou, Q.~Wang, and X.~Hu, ``Retiring dp: New distribution-level metrics for demographic parity,'' \emph{Transactions on Machine Learning Research}, 2023.

\bibitem{papadaki2022minimax}
A.~Papadaki, N.~Martinez, M.~Bertran, G.~Sapiro, and M.~Rodrigues, ``Minimax demographic group fairness in federated learning,'' in \emph{Proceedings of the 2022 ACM Conference on Fairness, Accountability, and Transparency}, 2022, pp. 142--159.

\bibitem{mougan2023demographic}
C.~Mougan, L.~State, A.~Ferrara, S.~Ruggieri, and S.~Staab, ``Demographic parity inspector: Fairness audits via the explanation space,'' \emph{arXiv preprint arXiv:2303.08040}, 2023.

\bibitem{de2023empirical}
A.~S. de~Oliveira, C.~Kaplan, K.~Mallat, and T.~Chakraborty, ``An empirical analysis of fairness notions under differential privacy,'' \emph{arXiv preprint arXiv:2302.02910}, 2023.

\bibitem{ferry2023addresing}
J.~Ferry, ``Addresing interpretability fairness \& privacy in machine learning through combinatorial optimization methods,'' Ph.D. dissertation, Universit{\'e} Paul Sabatier-Toulouse III, 2023.

\bibitem{wang2022towards}
A.~Wang, V.~V. Ramaswamy, and O.~Russakovsky, ``Towards intersectionality in machine learning: Including more identities, handling underrepresentation, and performing evaluation,'' in \emph{Proceedings of the 2022 ACM Conference on Fairness, Accountability, and Transparency}, 2022, pp. 336--349.

\bibitem{sattigeri2022fair}
P.~Sattigeri, S.~Ghosh, I.~Padhi, P.~Dognin, and K.~R. Varshney, ``Fair infinitesimal jackknife: Mitigating the influence of biased training data points without refitting,'' \emph{Advances in Neural Information Processing Systems}, vol.~35, pp. 35\,894--35\,906, 2022.

\bibitem{gardner2022subgroup}
J.~Gardner, Z.~Popovic, and L.~Schmidt, ``Subgroup robustness grows on trees: An empirical baseline investigation,'' \emph{Advances in Neural Information Processing Systems}, vol.~35, pp. 9939--9954, 2022.

\bibitem{ferry2023exploiting}
J.~Ferry, U.~A{\"\i}vodji, S.~Gambs, M.-J. Huguet, and M.~Siala, ``Exploiting fairness to enhance sensitive attributes reconstruction,'' in \emph{2023 IEEE Conference on Secure and Trustworthy Machine Learning (SaTML)}.\hskip 1em plus 0.5em minus 0.4em\relax IEEE, 2023, pp. 18--41.

\bibitem{cruz2023unprocessing}
A.~F. Cruz and M.~Hardt, ``Unprocessing seven years of algorithmic fairness,'' \emph{arXiv preprint arXiv:2306.07261}, 2023.

\bibitem{alvarez2023domain}
J.~M. Alvarez, K.~M. Scott, B.~Berendt, and S.~Ruggieri, ``Domain adaptive decision trees: Implications for accuracy and fairness,'' in \emph{Proceedings of the 2023 ACM Conference on Fairness, Accountability, and Transparency}, 2023, pp. 423--433.

\bibitem{cruz2022fairgbm}
A.~F. Cruz, C.~Bel{\'e}m, S.~Jesus, J.~Bravo, P.~Saleiro, and P.~Bizarro, ``Fairgbm: Gradient boosting with fairness constraints,'' \emph{arXiv preprint arXiv:2209.07850}, 2022.

\bibitem{bharti2024estimating}
B.~Bharti, P.~Yi, and J.~Sulam, ``Estimating and controlling for equalized odds via sensitive attribute predictors,'' \emph{Advances in Neural Information Processing Systems}, vol.~36, 2024.

\bibitem{simson2023using}
J.~Simson, F.~Pfisterer, and C.~Kern, ``Using multiverse analysis to evaluate the influence of model design decisions on algorithmic fairness,'' in \emph{HHAI 2023: Augmenting Human Intellect}.\hskip 1em plus 0.5em minus 0.4em\relax IOS Press, 2023, pp. 382--384.

\bibitem{nguyen23fix}
G.~Nguyen, S.~Biswas, and H.~Rajan, ``Fix fairness, don't ruin accuracy: Performance aware fairness repair using automl,'' in \emph{ESEC/FSE'2023: The 31st ACM Joint European Software Engineering Conference and Symposium on the Foundations of Software Engineering}, December 3-9, 2023 2023.

\bibitem{andreeva2004impact}
G.~Andreeva, J.~Ansell, and J.~Crook, ``Impact of anti-discrimination laws on credit scoring,'' \emph{Journal of Financial Services Marketing}, vol.~9, pp. 22--33, 2004.

\bibitem{chouldechova2018frontiers}
A.~Chouldechova and A.~Roth, ``The frontiers of fairness in machine learning,'' \emph{arXiv preprint arXiv:1810.08810}, 2018.

\bibitem{tizpaz2022fairness}
S.~Tizpaz-Niari, A.~Kumar, G.~Tan, and A.~Trivedi, ``Fairness-aware configuration of machine learning libraries,'' in \emph{Proceedings of the 44th International Conference on Software Engineering}, 2022, pp. 909--920.

\bibitem{chang2021privacy}
H.~Chang and R.~Shokri, ``On the privacy risks of algorithmic fairness,'' in \emph{2021 IEEE European Symposium on Security and Privacy (EuroS\&P)}.\hskip 1em plus 0.5em minus 0.4em\relax IEEE, 2021, pp. 292--303.

\bibitem{corbett2018measure}
S.~Corbett-Davies and S.~Goel, ``The measure and mismeasure of fairness: A critical review of fair machine learning,'' \emph{arXiv preprint arXiv:1808.00023}, 2018.

\bibitem{chen2024fairness}
Z.~Chen, J.~M. Zhang, F.~Sarro, and M.~Harman, ``Fairness improvement with multiple protected attributes: How far are we?''\hskip 1em plus 0.5em minus 0.4em\relax IEEE/ACM, 2024.

\bibitem{salewski2023context}
L.~Salewski, S.~Alaniz, I.~Rio-Torto, E.~Schulz, and Z.~Akata, ``In-context impersonation reveals large language models' strengths and biases,'' \emph{arXiv preprint arXiv:2305.14930}, 2023.

\bibitem{wang2023large}
P.~Wang, L.~Li, L.~Chen, D.~Zhu, B.~Lin, Y.~Cao, Q.~Liu, T.~Liu, and Z.~Sui, ``Large language models are not fair evaluators,'' \emph{arXiv preprint arXiv:2305.17926}, 2023.

\bibitem{yu2023large}
Y.~Yu, Y.~Zhuang, J.~Zhang, Y.~Meng, A.~Ratner, R.~Krishna, J.~Shen, and C.~Zhang, ``Large language model as attributed training data generator: A tale of diversity and bias,'' \emph{arXiv preprint arXiv:2306.15895}, 2023.

\bibitem{hernandez2019bargaining}
M.~Hernandez, D.~R. Avery, S.~D. Volpone, and C.~R. Kaiser, ``Bargaining while black: The role of race in salary negotiations.'' \emph{Journal of Applied Psychology}, vol. 104, no.~4, p. 581, 2019.

\bibitem{arceo2022gender}
E.~O. Arceo-Gomez, R.~M. Campos-Vazquez, R.~Y. Badillo, and S.~Lopez-Araiza, ``Gender stereotypes in job advertisements: What do they imply for the gender salary gap?'' \emph{Journal of Labor Research}, vol.~43, no.~1, pp. 65--102, 2022.

\bibitem{taylor2020salary}
L.~L. Taylor, J.~N. Lahey, M.~I. Beck, and J.~E. Froyd, ``How to do a salary equity study: With an illustrative example from higher education,'' \emph{Public personnel management}, vol.~49, no.~1, pp. 57--82, 2020.

\bibitem{platteau2021cognitive}
J.-P. Platteau and D.~U. Ontiveros, ``Cognitive bias in insurance: evidence from a health scheme in india,'' \emph{World Development}, vol. 144, p. 105498, 2021.

\bibitem{adult_income}
``Adult income dataset,'' \url{www.kaggle.com/datasets/wenruliu/adult-income-dataset}, 2023, accessed on August 1, 2023.

\bibitem{employee}
``Employee dataset,'' \url{www.kaggle.com/datasets/tawfikelmetwally/employee-dataset}, 2023, accessed on August 1, 2023.

\bibitem{insurance}
``Us health insurance dataset,'' \url{www.kaggle.com/datasets/teertha/ushealthinsurancedataset}, 2023, accessed on August 1, 2023.

\bibitem{Mehrabi2019ASO}
\BIBentryALTinterwordspacing
N.~Mehrabi, F.~Morstatter, N.~A. Saxena, K.~Lerman, and A.~G. Galstyan, ``A survey on bias and fairness in machine learning,'' \emph{ACM Computing Surveys (CSUR)}, vol.~54, pp. 1 -- 35, 2019. [Online]. Available: \url{https://api.semanticscholar.org/CorpusID:201666566}
\BIBentrySTDinterwordspacing

\bibitem{Nadeem2020StereoSetMS}
M.~Nadeem, A.~Bethke, and S.~Reddy, ``Stereoset: Measuring stereotypical bias in pretrained language models,'' in \emph{Annual Meeting of the Association for Computational Linguistics}, 2020.

\bibitem{chen2023fairness}
Z.~Chen, J.~Zhang, F.~Sarro, and M.~Harman, ``Fairness improvement with multiple protected attributes: How far are we?'' in \emph{46th International Conference on Software Engineering (ICSE 2024)}.\hskip 1em plus 0.5em minus 0.4em\relax ACM, 2023.

\bibitem{dutta2020there}
S.~Dutta, D.~Wei, H.~Yueksel, P.-Y. Chen, S.~Liu, and K.~Varshney, ``Is there a trade-off between fairness and accuracy? a perspective using mismatched hypothesis testing,'' in \emph{International conference on machine learning}.\hskip 1em plus 0.5em minus 0.4em\relax PMLR, 2020, pp. 2803--2813.

\bibitem{barlas2021see}
P.~Barlas, K.~Kyriakou, O.~Guest, S.~Kleanthous, and J.~Otterbacher, ``To" see" is to stereotype: Image tagging algorithms, gender recognition, and the accuracy-fairness trade-off,'' \emph{Proceedings of the ACM on Human-Computer Interaction}, vol.~4, no. CSCW3, pp. 1--31, 2021.

\bibitem{chen2022maat}
Z.~Chen, J.~M. Zhang, F.~Sarro, and M.~Harman, ``Maat: a novel ensemble approach to addressing fairness and performance bugs for machine learning software,'' in \emph{Proceedings of the 30th ACM Joint European Software Engineering Conference and Symposium on the Foundations of Software Engineering}, 2022, pp. 1122--1134.

\bibitem{cooper2021emergent}
A.~F. Cooper, E.~Abrams, and N.~Na, ``Emergent unfairness in algorithmic fairness-accuracy trade-off research,'' in \emph{Proceedings of the 2021 AAAI/ACM Conference on AI, Ethics, and Society}, 2021, pp. 46--54.

\bibitem{liu2022accuracy}
S.~Liu and L.~N. Vicente, ``Accuracy and fairness trade-offs in machine learning: A stochastic multi-objective approach,'' \emph{Computational Management Science}, vol.~19, no.~3, pp. 513--537, 2022.

\bibitem{Guo2020GraphCodeBERTPC}
D.~Guo, S.~Ren, S.~Lu, Z.~Feng, D.~Tang, S.~Liu, L.~Zhou, N.~Duan, J.~Yin, D.~Jiang, and M.~Zhou, ``Graphcodebert: Pre-training code representations with data flow,'' \emph{ArXiv}, vol. abs/2009.08366, 2020.

\bibitem{codebert}
\BIBentryALTinterwordspacing
Z.~Feng, D.~Guo, D.~Tang, N.~Duan, X.~Feng, M.~Gong, L.~Shou, B.~Qin, T.~Liu, D.~Jiang, and M.~Zhou, ``{C}ode{BERT}: A pre-trained model for programming and natural languages,'' in \emph{Findings of the Association for Computational Linguistics: EMNLP 2020}.\hskip 1em plus 0.5em minus 0.4em\relax Online: Association for Computational Linguistics, Nov. 2020, pp. 1536--1547. [Online]. Available: \url{https://aclanthology.org/2020.findings-emnlp.139}
\BIBentrySTDinterwordspacing

\bibitem{Ahmad2021UnifiedPF}
\BIBentryALTinterwordspacing
W.~U. Ahmad, S.~Chakraborty, B.~Ray, and K.-W. Chang, ``Unified pre-training for program understanding and generation,'' \emph{ArXiv}, vol. abs/2103.06333, 2021. [Online]. Available: \url{https://api.semanticscholar.org/CorpusID:232185260}
\BIBentrySTDinterwordspacing

\bibitem{CERT}
D.~Zan, B.~Chen, D.~Yang, Z.~Lin, M.~Kim, B.~Guan, Y.~Wang, W.~Chen, and J.-G. Lou, ``{CERT}: Continual pre-training on sketches for library-oriented code generation,'' in \emph{The 2022 International Joint Conference on Artificial Intelligence}, 2022.

\bibitem{bleu}
\BIBentryALTinterwordspacing
K.~Papineni, S.~Roukos, T.~Ward, and W.-J. Zhu, ``{B}leu: a method for automatic evaluation of machine translation,'' in \emph{Proceedings of the 40th Annual Meeting of the Association for Computational Linguistics}.\hskip 1em plus 0.5em minus 0.4em\relax Philadelphia, Pennsylvania, USA: Association for Computational Linguistics, Jul. 2002, pp. 311--318. [Online]. Available: \url{https://aclanthology.org/P02-1040}
\BIBentrySTDinterwordspacing

\bibitem{rouge}
\BIBentryALTinterwordspacing
C.-Y. Lin, ``{ROUGE}: A package for automatic evaluation of summaries,'' in \emph{Text Summarization Branches Out}.\hskip 1em plus 0.5em minus 0.4em\relax Barcelona, Spain: Association for Computational Linguistics, Jul. 2004, pp. 74--81. [Online]. Available: \url{https://aclanthology.org/W04-1013}
\BIBentrySTDinterwordspacing

\bibitem{Ren2020CodeBLEUAM}
\BIBentryALTinterwordspacing
S.~Ren, D.~Guo, S.~Lu, L.~Zhou, S.~Liu, D.~Tang, M.~Zhou, A.~Blanco, and S.~Ma, ``Codebleu: a method for automatic evaluation of code synthesis,'' \emph{ArXiv}, vol. abs/2009.10297, 2020. [Online]. Available: \url{https://api.semanticscholar.org/CorpusID:221836101}
\BIBentrySTDinterwordspacing

\bibitem{Evtikhiev2022OutOT}
\BIBentryALTinterwordspacing
M.~Evtikhiev, E.~Bogomolov, Y.~Sokolov, and T.~Bryksin, ``Out of the bleu: how should we assess quality of the code generation models?'' \emph{J. Syst. Softw.}, vol. 203, p. 111741, 2022. [Online]. Available: \url{https://api.semanticscholar.org/CorpusID:251371647}
\BIBentrySTDinterwordspacing

\bibitem{huang2024effibench}
D.~Huang, Y.~Qing, W.~Shang, H.~Cui, and J.~M. Zhang, ``Effibench: Benchmarking the efficiency of automatically generated code,'' \emph{arXiv preprint arXiv:2402.02037}, 2024.

\bibitem{ling2024evaluating}
L.~Ling, ``Evaluating social bias in code generation models,'' in \emph{Companion Proceedings of the 32nd ACM International Conference on the Foundations of Software Engineering}, 2024, pp. 695--697.

\end{thebibliography}

% \appendix

% \section{Appendix}

\end{document}